\documentclass[twocolumn,pra,showpacs]{revtex4}
\usepackage{graphicx}

\begin{document}
\title {Fragmentation of Bose-Einstein Condensates}
\author{Erich J.  Mueller,$^a$ Tin-Lun Ho,$^b$, Masahito Ueda,$^c$ and
Gordon Baym$^d$}
\affiliation{$^a$Laboratory for Atomic and Solid State Physics, Cornell
University,
  Ithaca, NY}
\affiliation{$^b$Department of Physics, The Ohio State University,
  Columbus, Ohio 43210}
\affiliation{$^c$Department of Physics, Tokyo Institute of Technology,
  2-12-1 Ookayama, Meguro-ku, Tokyo 152-8551, Japan}
\affiliation{$^d$Department of Physics, University of Illinois at
Urbana-Champaign, 1110 W, Green St.,  Urbana IL 61801}

\begin{abstract}

    We present the theory of bosonic systems with multiple condensates,
unifying disparate models which are found in the literature, and discuss how
degeneracies, interactions, and symmetries conspire to give rise to this
unusual behavior.  We show that as degeneracies multiply, so do the types of
fragmentation, eventually leading to strongly correlated states with no trace
of condensation.

\end{abstract} \pacs{03.75.Fi}

\maketitle

\section {Bose-Einstein condensation, fragmented condensates, and strongly
correlated states }

    Bose-Einstein condensation (BEC) is a very robust phenomenon.  Because of
Bose statistics, non-interacting bosons seek out the lowest single particle
energy state and (below a critical temperature $T_{c}$) condense into it, even
though many almost degenerate states may be nearby.  At temperature $T=0$, the
condensate contains all the particles in the system \cite{huang}.  Although
this remarkable phenomenon was originally predicted by Einstein for
non-interacting systems \cite{einstein}, it became understood, starting with the work of F.
London \cite{london}, that it occurs in strongly interacting systems such as
$^4$He.  While interactions remove particles from the condensate into other
states, the energy gain by macroscopically occupying the lowest energy state
(or another state, such as a vortex) is sufficiently great that the interactions
in a Bose condensed system manage only to deplete a fraction of the
condensate, but not destroy it \cite{bogoliubov}.  This is the case for liquid
$^4$He \cite{Penrose1956a} as well as dilute gases of bosonic alkali atoms
\cite{boseatoms}.

    In certain situations, however, a system does not condense into a {\em
single} condensate \cite{nsj,Nozieres1995a,Wilkin1998a,girardeau,Koashi,Ho1999a,
castin,peth,Rokhsar1998a,spekkens,jav,dukelsky}.  In this paper we explore the
physics of condensation when the ground state can contain several condensates
-- situations of {\em fragmented} condensation.  This possibility arises
naturally from the very concept of BEC -- i.e., that the {\em non-degenerate}
ground state of a single particle Hamiltonian $h_0$ is macroscopically
occupied.  How, instead, does condensation take place when the ground state of
$h_0$ is degenerate, with two or more states competing simultaneously for
condensation?  What happens if the ground state degeneracy $G$ is not just of
order unity, but of order $N$, the number of particles?  And what happens if
$G$ becomes much greater than $N$, or even approaches infinity?  How do the
bosons distribute themselves in these competing levels?  In all these cases,
interaction effects play an important role in determining the structure of the
many-body ground state.  Different types of interactions produce different
fluctuations (such as those of phase, number, or spin) and lead to different
classes of ground states.

    Exact degeneracy is difficult to achieve, since small external fields or
weak tunneling effects are generically present.  Such energy-splitting effects
lead to a non-degenerate ground state and therefore favor the formation of a
single condensate.  Only when interaction effects dominate over energy
splittings can a single condensate break up.

    Considerations of the effect of ground state degeneracies are not merely
theoretical exercises.  Rather, such degenracies occur in a wide range of
current experiments in cold atoms.  The cases where the degeneracy $G$ is of
order unity are related to bosons with internal degrees of freedom.  Examples
include a pseudospin-1/2 Bose gas made up, e.g., of two spin states $|F=1,
m=-1\rangle$ and $|F=2, m=1\rangle$) of $^{87}$Rb \cite{jilaspinexp}, and a
spin-1 or spin-2 Bose gas such as $^{23}$Na or $^{87}$Rb in an optical trap
\cite{spinexperiments}.  In the former case, even though the two spin states
of $^{87}$Rb are separated by a hyperfine splitting of order GHz, they can be
brought to near degeneracy, with $G=2$, by applying an external rf field.  In
the spin-1 Bose gas, the three spin states are degenerate at zero external
magnetic field, and $G=3$.  The case of $G\sim N^{1/2}$ is encountered in one
dimensional geometries \cite{1dexp}, where the density of states has a power
law singularity at low energies.  The case of $G\sim N$ is realized for bosons
in optical lattices \cite{mott} with $G$ sites each having a few bosons; in
the limit of zero tunneling, $G$ equivalent sites compete for bosons.  The
case $G>>N$ is realized in rotating Bose gases with very large angular
momentum, $L$, in a transverse harmonic trap \cite{hall,wg}.  As $L$
increases, the rotation frequency $\Omega$ of the atom cloud approaches the
frequency of the transverse harmonic trap, $\omega_{T}$, causing the single
particle states to organize into Landau orbitals, which become infinitely
degenerate as $\Omega \to \omega_{T}$.  The great diversity of phenomena in
these experiments is a manifestation of the physics of Bose condensation for
varying degrees of degeneracy.

    As we shall see, when the degeneracy is low ($G\sim 1$), a single
condensate can break up into $G$ condensates.  Following Nozi\`eres
\cite{nsj,Nozieres1995a}, we refer to such a state as a {\it fragmented
condensate} (define more precisely below).  We caution, however, that there
are many sorts of fragmented states.  Even degeneracy as small as $G=2$ can
give rise to fragmented condensates with distinct properties which depend on
whether interaction effects favor phase or number fluctuations.  System with
larger degeneracies usually have a larger number of relevant
 interaction parameters and are more
easily influenced by external fields.  As a result, they have a greater
variety of fragmented states.  For degeneracies $G$ comparable to or much
greater than the particle number, the system does not have enough particles to
establish separate condensates.  Instead, interactions effects tend to
distribute bosons among different degenerate single particle states in a
coherent way, establishing correlations between them.  In such {\it strongly
correlated} systems, interaction effects obliterate all traces of a
conventional singly condensed state.

    We stress at the outset that while interaction effects can cause
fragmentation in the presence of degenerate single particle states, the
presence of near degeneracies does not force the condensate to fragment.  In
many cases the ground state for a {\em macroscopic} Bose system is a
conventional single condensate.  Creating a fragmented state typically
requires carefully tuning the parameters of the system, and whether such a
state can occur in practice is very much dependent on the system.  In optical
lattices and double well systems, where the tunneling between wells can be
tuned arbitrarily finely, a fragmented state can easily be achieved.  Yet in
other systems such as a spin-1 Bose gas in a single trap, or a rotating Bose
gas, the parameter range allowing the existence of fragmented states scales
like $1/N$, making it difficult to realize these ground states unless the
number of particles is reduced to $\sim 10^{3}$ or fewer.

    In this paper, we focus on fragmentation in the ground state of Bose
systems.  We consider a number of canonical examples (double-well systems,
spin-1 Bose gases, and rotating Bose gases), which have increasing degeneracy
in the single particle Hamiltonian.  These examples illustrate the origin of
fragmentation, the variety of fragmented states, and their key properties.  We
will not discuss here fragmentation in optical lattices or in dynamical
processes, for they are a sufficiently large subject to require separate
discussions.  To begin, we first define condensate fragmentation more
precisely, and the discuss general properties of certain classes of fragmented
states.

\section{Definitions of condensation and fragmentation}

    The concept of Bose-Einstein condensation was generalized to interacting
systems by Penrose and Onsager\cite{Penrose1951a,Penrose1956a} in the 1950s,
by defining condensation in terms of the single particle density matrix,
\begin{equation}
    \rho^{(1)}({\bf r},{\bf r}^\prime)=\langle\psi^\dagger({\bf
    r^\prime})\psi({\bf r})\rangle,
  \label{rho}
\end{equation}
where $\psi^{\dagger}({\bf r})$ creates a scalar boson at position ${\bf
r}$, and $\langle \rangle$ is the thermal average at temperature $T$.  Since
$\rho^{(1)}$ is a Hermitian matrix with indices ${\bf r}$ and ${\bf r'}$,
it can be diagonalized as,
\begin{equation}
   \rho^{(1)}({\bf r},{\bf r}^\prime)=\sum_{i} N_{i}(T) f_{i}({\bf r}')^{\ast}
   f_{i}({\bf r}),
 \label{rho1}
\end{equation}
where the $N_{i}(T)$ are the eigenvalues and $f_{i}({\bf r})$ the
orthonormal eigenfunctions of $\rho^{(1)}$; $\int f^{\ast}_{i}({\bf
r})f_{j}^{}({\bf r})= \delta_{ij}$.  Setting ${\bf r}={\bf r}^\prime$ and
integrating over ${\bf r}$ we have $\sum_{i}N_{i}(T)=N$, where $N$ is the
number of particles.  We label the eigenvalues in descending order, i.e.,
$N_{0}\geq N_{1} \geq N_{2} \geq \cdots$.  Equation (\ref{rho1}) implies that
if one measures the number of bosons in the single particle state $f_i$, one
finds $N_i$.  This does not mean that the wavefunction of the many particle
interacting system is a product of such single particle eigenstates.  Unless
there are special reasons (such as strict symmetry constraints, e.g.,
translational invariance), the eigenfunctions $\{f_i\}$ need not be the same
as the single particle eigenstates $\{u_i\}$ of the single particle
Hamiltonian of the non-interacting system.  Two relevant examples are a
homogeneous system of particles in free space, and a system in a harmonic
trap.  In the former case, where the momentum ${\bf k}$ is a good quantum
number, we have $i={\bf k}$, and $\{ f_{i}\}$ and $\{ u_{i}\}$ are the same
plane-wave momentum eigenstates.  In an inhomogeneous trapped system, there
is no simple relation between the $\{f_i\}$ and the $\{u_j\}$ \cite{dubois}.

    The usual situation of Bose condensation corresponds to the one eigenvalue
$N_0$ being of order $N$, while other eigenvalues are of order unity, i.e.,
\begin{eqnarray}
 {\rm BEC:} \,\,\,\,\,\,\,\,\,\,
 \rho^{(1)}({\bf r},{\bf r}^\prime) &=&
 N_{0}(T) f_{0}({\bf r}')^* f_{0}({\bf r}) + \cdots\\
 \label{BEC}
 &\equiv& \Psi^{(0)*}({\bf r}')\Psi^{(0)}({\bf r}) + \cdots\, ,
  \label{scalarBEC}
\end{eqnarray}
or simply
\begin{equation}
  \rho^{(1)} =  N_0 |\Psi^{(0)}\rangle \langle \Psi^{(0)}|+ \cdots\, ,
\end{equation}
where $\Psi^{(0)}({\bf r})\equiv \sqrt{N_{0}} f_{0} ({\bf r})$, and
$(\cdots)$ denotes terms with eigenvalues $N_{i}\sim {\cal O}(1)$.  Since the
macroscopic term in Eq.~(\ref{BEC}) is identical to the density matrix of the
pure single particle quantum state, $|\Psi^{(0)}\rangle$, the function
$\Psi^{(0)}({\bf r})$ is often referred to as the ``macroscopic wavefunction"
of the system.  Systems in which $\rho^{(1)}$ has only one macroscopic
eigenvalue, Eq.~(\ref{BEC}), have {\em single condensates}.  The advantage of
the Penrose-Onsager characterization of BEC, Eq.~(\ref{BEC}), is that it
applies to both interacting and non-interacting systems, since it makes no
reference to dynamics.  Penrose and Onsager also demonstrated the remarkable
fact that Eq.~(\ref{BEC}) holds for a Jastrow function, which is a reasonable
approximation to the ground state of a system of hard core bosons, therefore
substantiating Eq.~(\ref{BEC}) as a general property of interacting Bose
systems.

    The Penrose-Onsager characterization can be easily generalized to bosons
with internal degrees of freedom, labeled by an index $\mu$.  With field
operator $\psi^{\dagger}_{\mu}({\bf r})$, the single particle density matrix
is
\begin{eqnarray}
 \rho^{(1)}({\bf r,\mu};{\bf r}^\prime,\mu^\prime) &=&
 \Psi^{(0)*}_{\mu^\prime}({\bf r}')\Psi^{(0)}_\mu({\bf r})
 + \cdots\, .
  \label{scalarBECmu}
\end{eqnarray}

    In a conventional noncondensed system, such as a zero temperature gas of
non-interacting fermions, or a high temperature gas of bosons, all the
occupation numbers are small:  $N_i\sim {\cal O}(N^{0})$.  A conventional
singly condensed system has one large eigenvalue $N_0\sim {\cal O}(N)$, with
all other eigenvalues small, $N_{i>0}\sim {\cal O}(N^{0})$.  A fragmented
system has $q>1$ large eigenvalues, $N_{i<q}\sim {\cal O}(N^{1})$.  There is
clearly a range of other possiblilites such as having an extremely large
number $q\sim {\cal O}(N^{1/2})$ eigenvalues, each of which are of size
$N_{i<q}\sim {\cal O}(N^{1/2})$.  This latter case occurs in an interacting
system of one dimensional bosons, and is associated with a phase incoherent
{\em quasicondensate} \cite{lowd,1dexp}.

\subsection{Simple example of fragmentation and its relation with
  coherent states}\label{noz}

    Before examining the origin of fragmentation in detail, let us consider a
basic example of fragmentation -- the Nozi\`eres model \cite{Nozieres1995a}.
Consider a system of $N$ bosons each of which have available two internal
states; 1 and 2.  As we consider in detail later, this model can also be used to describe atoms in a double well potential.
The Hamiltonian of  Nozi\`eres' model consists solely of an interaction between atoms in the two spin states,
\begin{eqnarray}
  H &=& \frac{g}{2}a^\dagger_1  a^\dagger_2 a_2 a_1
   = \frac{g}{2}n_1 n_2,
  \label{nozieres}
\end{eqnarray}
where the $a^{\dagger}_i$ create bosons in state $i=1,2$, and
$n_i=a^{\dagger}_i a_i$ is the number of particles in $i$.  The interaction
between the particles can be either repulsive ($g>0$) or
attractive ($g<0$).  The eigenstates have a definite number of particles in
each well $|N_1,N_2\rangle$, with $N=N_1+N_2$, and energy
\begin{equation}
   E = \frac{g}{2}N_1N_2.
\end{equation}
Clearly, for $g>0$, the ground state is two-fold degenerate, with $N_1=N,
N_2=0$, or $N_1=0, N_2=N$; these states have single condensates, whose density
matrices have eigenvalues 0 and $N$.  On the other hand, for $g<0$, the
state with $N_1=N_2=N/2\,$ has lowest energy.  This {\em Fock}-state,
\begin{equation}
  |F\rangle = \frac{ a^{\dagger N/2}_{1}a^{\dagger N/2}_{2}}{(N/2)!}
  |0\rangle,
 \label{F}
\end{equation}
has a fragmented condensate; the corresponding single particle density
matrix,
\begin{equation}
   \rho^{(1)} = \langle a^{\dagger}_{\mu} a^{}_{\nu}\rangle = \frac{N}{2}
  \left(\begin{array}{cc} 1 & 0 \\ 0
   &1\end{array}\right),
   \label{rhoF}
\end{equation}
has two macroscopic eigenvalues.

    One can contrast this fragmented state to the coherent state
\begin{equation}
 |\phi_N\rangle = \frac{1}{\sqrt{2^N N!}}\left(e^{-i\phi/2}
   a^\dagger_1 + e^{i\phi/2} a^\dagger_2\right)^N|0\rangle
  \label{Cphi}
\end{equation}
in which $N$ bosons are condensed into the single particle state
$(e^{-i\phi/2} a^{\dagger}_{1} + e^{i\phi/2} a^{\dagger}_{2} )/\sqrt{2}$.  The
coherent state is an example of a single condensate, where the single particle
density matrix, $\rho^{(1)}= \langle a^{\dagger}_{\mu} a^{}_{\nu}\rangle$, is
\begin{eqnarray}
  \rho^{(1)} &=& \frac{N}{2} \left(\begin{array}{cc} 1 &
  e^{i\phi} \\ e^{-i\phi} &1\end{array}\right)\\ &=&
  \frac{N}{2} \left(\begin{array}{c} e^{-i\phi/2} \\
  e^{i\phi/2}\end{array}\right)^{\ast} \begin{array}{cc}
  (e^{-i\phi/2} & e^{i\phi/2} ) \\ & \end{array}.
  \label{rhoC}
\end{eqnarray}

    The difference between Eqs.~(\ref{rhoC}) and (\ref{rhoF}) is the absence
of the off-diagonal term $\langle a^{\dagger}_{1}a^{}_{2}\rangle$ in the
latter, which represents the loss of phase coherence in state (pseudospin)
space.

\subsection{Relation between Fock and coherent states}

    The Fock state (\ref{F}) is an average over all coherent phase states
$|\phi_N\rangle$, as we see from the relation,
\begin{eqnarray}
  |F\rangle &=& \frac{2^{N/2}}{\sqrt{N!}}\left(N/2\right)! \int^{\pi}_{-\pi}
  \frac{ {\rm d}\phi}{2\pi}|\phi_N\rangle \\&\approx&  \left(\frac{\pi
    N}{2}\right)^{1/4} \int^{\pi}_{-\pi}
  \frac{ {\rm d}\phi}{2\pi}|\phi_N\rangle;
  \label{F-C}
\end{eqnarray}
the latter relation holds for $N>>1$.  As we discuss in the next section,
this connection is very useful for understanding the origin of various ground
states.  An important implication of this relation is that {\em for a
macroscopic system, the expectation value of any $p$-body operator},
$\hat{O_p} \sim a^{\dagger}_{\mu_{1}}a^{\dagger}_{\mu_{2}} \dots
a^{\dagger}_{\mu_{p}} a^{}_{\nu_{p}} \dots a^{}_{\nu_{2}}a^{}_{\nu_{1}}$, {\em
in the Fock state is indistinguishable from that in an ensemble of coherent
phase states $|\phi\rangle$, as long as $p\ll N$},
\begin{equation}
 \langle F| \hat{O_p}|F\rangle = \int^{\pi}_{-\pi} \frac{ {\rm d}\phi}{2\pi}\,
 \langle \phi_N|\hat{O_p}|\phi_N \rangle.
 \label{F=C}
\end{equation}

    This equation, which we shall prove momentarily, shows that by measuring
quantities associated with few body operators, one cannot distinguish a Fock
state from an ensemble of coherent states with random phases
\cite{Javmeasure,castindalibard}.  An illustration of this effect is the
interference of two condensates initially well separated from each other.
Prior to any measurement process, the system is in a Fock state, Eq.(\ref{F}),
since there is no phase relation between the two condensates.  Experimentally,
in any single shot measurement (a photo of the interfering region), one finds
an interference pattern consisting of parallel fringes whose location is
specified by a phase $\phi$, as if the two far away condensates actually had a
well defined relative phase with this value \cite{mitinterference}.  The value
of $\phi$, however, varies randomly from shot to shot, so that if one
averages over all the measurements, the interference fringes average out, as
described simply by Eq.  (\ref{F=C}) \cite{puzzle}.

    To prove Eq.~(\ref{F=C}), we use Eq.~(\ref{F-C}) to write,
\begin{equation}
 \langle F| \hat{O_p}|F\rangle = \sqrt{\frac{\pi N}{2}}\int^{\pi}_{-\pi}
 \frac{ {\rm d}\phi'}{2\pi}\frac{ {\rm d}\phi}{2\pi}\,
 \langle \phi'_N|\hat{O_p}|\phi_N \rangle;
\end{equation}
it is then sufficient to show that $\langle
\phi'_{N}|\hat{O}_{p}|\phi_{N}\rangle$ vanishes unless the phases $\phi$ and
$\phi'$ are very close to each other.  We note that
\begin{eqnarray}
  \langle \phi'_N| \phi_N\rangle  & =
 \cos^N\left((\phi-\phi')/2\right)
 \approx e^{-N(\phi-\phi')^2/8} \nonumber \\
 & \approx \sqrt{8\pi/N} \delta(\phi-\phi'),
 \label{phi-phi'}
\end{eqnarray}
for $N>>1$.  For $p$-body operators of the form $\hat{O}_{p} =
{a^\dagger_1}^{q'} {a^\dagger_2}^{p-q'} {a_2}^{p-q}{a_1}^q$,
\begin{eqnarray}
  \langle \phi'_N| \hat{O}_{p} | \phi_N\rangle &=&
   \nonumber\ \\  && \hspace{-66pt}\frac{N!}{2^p(N-p)!}
   e^{i[\phi(p/2-q)-\phi'(p/2-q')]} \langle \phi'_{N-p}| \phi_{N-p}\rangle.
\end{eqnarray}
For $N>>p$, we see from Eq.~(\ref{phi-phi'}) that $\langle \phi'_N|
\hat{O}_{p} |\phi_N\rangle = \sqrt{8\pi/N} \langle \phi_N| \hat{O}_{p} |
\phi_N\rangle \delta(\phi - \phi') + {\cal O}(1/N)$, from which
Eq.~(\ref{F=C}) follows.

\section{Characteristic examples of fragmentation}\label{ex}

    We now develop three different examples that illustrate the origin of
fragmentation.  These examples are chosen to illustrate the increasingly complex
behavior of fragmentation when the number of degenerate single particle states increases.

\subsection{Scalar bosons in double well}\label{twowellsec}

    After the model of sec.~\ref{noz}, the simplest model with a fragmented
ground state is that of bosons in a double well potential with tunneling
between the wells.  Unlike in the previous example, this model produces two
distinct types of fragmented states.  We label the wells by $i=1,2$; we assume
that there is only one relevant state in each well, and that particles within
a given well have an interaction, $U$, which can be either repulsive ($U>0$)
or attractive ($U<0$).  We take the Hamiltonian to be
\begin{equation}
  H = - t(a_1^{\dagger}a_{2} + a_{2}^{\dagger}a^{}_{1} ) +
  \frac{U}{2}\left[ n_{1}(n_{1}-1) + n_{2}(n_{2}-1) \right],
  \label{2site}
\end{equation}
where the $a^{\dagger}_i$ creates a boson in well $i$, and
$n_i=a^{\dagger}_i a_i$ is the number of particles in well $i$.  The first
term describes tunneling between the wells via a tunneling matrix element $t$
(which we assume to be real and positive).  The form $Un_{i}(n_{i}-1)/2=
Ua^{\dagger}_{i}a^{\dagger}_{i}a_{i}a_{i}/2$ is the usual contact interaction
$(g/2)\int \psi^{\dagger}\psi^{\dagger}\psi\psi$ reduced to the single mode in
each well.  For fixed number of particles $n_{1}+ n_{2} = N$, the interaction
term can be written simply as
\begin{equation}
 \hat{U} = \frac{U}{4}\left[ \left(n_{1} - n_{2}\right)^2  +
     N^2- 2N\right]
\label{U} \end{equation}
This model, simple as it is, has wide applicability to many physical
situations:  atoms in a double well potential \cite{dwexp}, internal hyperfine
states coupled by electromagnetic fields \cite{jilaspinexp,spinexperiments},
atoms in a rotating toroidal trap \cite{uedaleggett}, or wave packets in an
optical lattice \cite{muellerswallowtails}.

    In solving this model it is useful to write Eq.~(\ref{2site}) in the
Wigner-Schwinger pseudospin representation \cite{lqm}.  We introduce the
operators
\begin{eqnarray}
  J_x &=& (a^{\dagger}_1a_2+ a^{\dagger}_2a_1)/2,   \,\,\,\,\, J_y =
  (a^{\dagger}_1a_2- a^{\dagger}_2a_1)/2i,
  \nonumber \\
   & & J_z = (a^{\dagger}_1a_1 - a^{\dagger}_2a_2)/2,
\end{eqnarray}
which obey the angular momentum commutation relation $[J_{i}, J_{j}]=
i\epsilon_{ijk}J_{k}$, $i =x,y,z$, and satisfy
\begin{equation}
  {\bf J}^2 = J_x^2 + J_y^2 + J_z^2 = \frac{N}{2}\left(\frac{N}{2}+1\right).
\end{equation}
The Hamiltonian (\ref{2site}) can be written in terms of ${\bf J}$ as
\begin{equation}
  H = - 2t J_{x} + U(J_{z}^2 + {\bf J}^2 -N).
   \label{spinH}
\end{equation}

\subsubsection{Mean-field solution}\label{dwmf}

    As we shall see, the Hamiltonian (\ref{2site}) can be solved exactly.
Nonetheless, the mean-field solutions illustrate much of the physics of the
true ground state.  They also allow one to see the kind of fluctuations about
the mean-field state that lead to condensate fragmentation.

    The mean-field states are of the form of (pseudo)spinor condensates
\begin{equation}
  |\theta, \phi \rangle = \frac1{\sqrt{N!}}(u a^{\dagger}_{1} + v
   a^{\dagger}_{2})^{N}|0\rangle,
\label{n}
\end{equation}
where $u = e^{-i\phi/2} \cos(\theta/2)$ and $v = e^{i\phi/2}
\sin(\theta/2)$.  The matrix elements of the density matrix in this state are
$\langle a^{\dagger}_1a_1\rangle = \langle n_1\rangle = N\cos^2(\theta/2)$,
$\langle a^{\dagger}_2a_2\rangle= \langle n_2\rangle = N\sin^2(\theta/2)$, and
$\langle a^{\dagger}_1 a_2\rangle=N\sin(\theta/2) \cos(\theta/2) e^{i\phi}$.
The angles $\theta$ and $\phi$ therefore characterize the density and phase
difference between the bosons in the two wells.  In the pseudospin language,
the state (\ref{n}) describes a ferromagnet with total spin, $\langle {\bf
J}\rangle =(N/2)\hat{\bf n}$, where $\hat{\bf n} = (\sin\theta \cos\phi,
\sin\theta \sin\phi,\cos\theta)$ is the unit vector with polar angles
$(\theta, \phi)$.  According to Eq.~(\ref{spinH}), its energy is
\begin{eqnarray}
 \label{Euv}
    E(\theta, \phi ) &=& \langle \theta, \phi |H| \theta, \phi
    \rangle\\\nonumber &=&
   -tN\cos\phi \sin\theta + U\left(\frac{N^2}{4} (\cos^2\theta +1)
   -\frac{N}{2}\right).
\end{eqnarray}

    For a repulsive interaction, $U>0$, $E(\theta, \phi)$ is minimum at
$\phi=0$, $\theta=\pi/2$; or $\hat{\bf n}= \hat{\bf x}$.  The mean-field
approach therefore selects the non-interacting ground state $|C\rangle =
(a^{\dagger}_{1}+ a^{\dagger}_2)^{N}|0\rangle/\sqrt{2^N N!}= |\hat{\bf
x}\rangle$ as optimal.  Condensates in the neighborhood of $|C\rangle$ have
energy
\begin{equation}
  E(\pi/2+\epsilon,\phi) = E(\pi/2,0)+ \frac12 tN\phi^2
  + \frac{N}{4}(2t + UN)\epsilon^2  +\dots.
  \label{d2mean}
\end{equation}
From this result one can begin to see problems with the mean-field
solution: as $t\to 0$ with fixed $U$, the energy of phase fluctuations
($\phi$) vanishes; therefore quantum fluctuations begin to mix in many nearly
degenerate phase states, $|\pi/2, \phi\rangle$.

    The mean-field solution for attractive $U$ is very different from that of
repulsive $U$.  The solution depends on whether $|U|N<2t$ or $|U|N>2t$.  In
the former case, $|C\rangle$ is locally stable since the energy of
fluctuations, described by Eq.~(\ref{d2mean}), remain positive definite.
However, the stiffness constant for density fluctuations ($\epsilon$) is lower
than that for phase fluctuations ($\phi$).  Thus, as $|U|$ increases, density
fluctuations become dominant.  Note, however, that the condition $|U|N<2t$
only occurs for a bounded number of particles, and under ordinary
circumstances is not expected to be achievable in macroscopic systems.  On the
other hand, when $|U|N>2t$, we see from Eq.~(\ref{Euv}) that the optimal
states satisfy $\sin\theta = 2t/|U|N$.  There are two degenerate solutions:
$\theta = \theta_0= \arcsin(2t/|U|N)$ and $\pi - \theta_0$.  For $|UN|/2t>>1$,
these two states approach $\sim a^{\dagger N}_1 |0\rangle/\sqrt{N!}$ and $\sim
a^{\dagger N}_2|0\rangle/\sqrt{N!}$, corresponding to all particles being in
well 1 or 2 respectively.

\subsubsection{Exact ground states}

    We now construct the exact ground state of the Hamiltonian, (\ref{2site}),
and see how interactions can cause condensate fragmentation.  At the same
time, we can see how different types of interaction cause different types
of fragmented states.

    {\em Non-interacting case:} Let us first consider the simplest case of
non-interacting bosons, with Hamiltonian $H = -t (a^{\dagger}_{1} a^{}_{2} +
{\rm h.c.})$.  The single particle eigenstates are the symmetric state $(a_{1} +
a_{2})/\sqrt{2}$ and antisymmetric state $(a_{1} - a_{2})/\sqrt{2}$ with
energy $-t$ and $t$ respectively.  For a system of $N$ bosons, the ground
state is
\begin{equation}
  |C\rangle  = \frac{1}{\sqrt{2^NN!} }
  (a^{\dagger}_1 + a^{\dagger}_2)^{N} |0\rangle
\label{C}
\end{equation}
with energy $-tN$.  The single particle
density matrix of this state is
\begin{equation}
  \langle a^{\dagger}_{\mu}a^{}_{\nu}\rangle_{_{C}} = \frac{N}{2}\left(
  \begin{array}{cc} 1& 1\\
  1 & 1\end{array}\right)
  = N\left( \begin{array}{c}  \frac{1}{\sqrt{2}}\\ \frac{1}{\sqrt{2}}
  \end{array}\right)
   \begin{array}{cc}\left(\frac{1}{\sqrt{2}}\right. , & \left.\frac{1}{\sqrt{2}} \right), \\ &
\end{array}
\label{spdm-coh}
\end{equation}
which has a single macroscopic eigenvalue $\lambda =N$.  The ground state
is therefore a single condensate with condensate wavefunction $\zeta^{T}_{\mu}
=\sqrt{N/2} (1, 1)$ (the superscript ``$T$" stands for transpose).

    Since the ground state $|C\rangle$ is a linear combination of number
states $|n_{1}, n_{2}\rangle = a^{\dagger n_{1}}_{1} a^{\dagger
n_{2}}_{2}|0\rangle/\sqrt{n_{1}! n_{2}!}$, the 
number of particles in each well fluctuates.

We calculate the number fluctuations
of the coherent state $|C\rangle$ by writing it in the number basis.
For even $N$, we have
\begin{equation}
  |C\rangle = \sum_{\ell = -N/2}^{N/2} \Psi^{(0)}_{\ell}  |\ell\rangle,
\end{equation}
where $|\ell\rangle \equiv  |\frac{N}{2}+\ell, \frac{N}{2}-\ell\rangle$,  and
\begin{equation}
  \Psi^{(0)}_{\ell}= \left(\frac{ N!}{ 2^{N}(\frac12 N+\ell)! (\frac12
   N-\ell)!}\right)^{1/2}
   \approx \frac{e^{-\ell^2/N} }{(\pi N/2)^{1/4}}.
 \label{Ao}
 \end{equation}
The number fluctuations are then
 \begin{equation}
  \langle \Delta N_{1}^2 \rangle=\langle(N_1-\langle N_1\rangle)^2\rangle
  \approx \int {\rm d} \ell \, \frac{\ell^2 e^{-2\ell^2/N} }{\sqrt{\pi N/2} }
  = N/4,
\end{equation}
which, despite the approximations made in this derivation, coincides with the exact result.

    {\em Interacting Case:} The many-body physics of this double well system
is completely tractable.  While one can calculate the properties the ground
state numerically to arbitrary accuracy, we derive below all the essential
features of the ground state by studying the effect of interactions on the
non-interacting ground state, i.e., the coherent state $|C\rangle$.  We shall
see that depending on whether the interactions are repulsive or attractive,
the coherent state can be turned into one of two distinct fragmented states:
a `Fock-like' state or a `Schr\"odinger cat-like' state.

    We first look at the Schr\"odinger equation of this system.  Writing
the ground state in the number basis,
\begin{equation}
|\Psi\rangle = \sum_{\ell = -N/2}^{N/2} \Psi_{\ell} |\ell\rangle,
\end{equation}
we can write the Schr\"odinger equation, $H|\Psi\rangle = E|\Psi\rangle$,
where $H$ is given by Eq.~(\ref{2site}), as
\begin{equation}
  E\Psi_{\ell} = - t_{\ell+1} \Psi_{\ell+1} - t_{\ell}\Psi_{\ell-1} + U\ell^2
  \Psi_{\ell},
\label{Sch}
\end{equation}
\begin{equation}
  t_{\ell} = t \sqrt{(N/2 + \ell)(N/2 - \ell + 1)}.
 \label{tell}
 \end{equation}
The many-body problem then reduces to a one-dimensional tight binding
model in a harmonic potential.  The special feature of this model is that the
tunneling matrix element $t_{\ell}$ is highly non-uniform \cite{spintrans}:
$t_{\ell}\sim N/2$ for $\ell \approx 0$, and $t_{\ell} \sim \sqrt{N/2}$ for
$\ell \approx \pm N/2$.  This non-uniformity is a consequence of bosonic
enhancement, $a^{\dagger}|N\rangle = \sqrt{N+1}|N+1\rangle$ and $a^{}|N\rangle
= \sqrt{N}|N-1\rangle$, which increases the matrix element by a factor of
$\sqrt{N}$ when removing a particle from a system with $N$ bosons.  As a
result, $\langle \ell +1 | a^{\dagger}_{1} a^{}_{2} |\ell\rangle$ is maximum
when both wells have equal number of bosons (i.e., $\ell \sim 0$), and drops
rapidly when the difference in boson numbers between the wells begins to
increase, (i.e., $\ell >>1$).  A consequence is that hopping favors
wavefunctions $\Psi_{\ell}$ having large amplitudes near $\ell =0$.  For
example, in the non-interacting case the wavefunction, Eq.~(\ref{Ao}), is a
sharply peaked Gaussian at $\ell=0$.

  The interaction term, $U(n_{1}-n_{2})^2/4$, leads to
a harmonic potential in Eq.~(\ref{Sch}).  Repulsive interactions $(U>0)$
suppress number fluctuations, meaning that the Gaussian distribution
[Eq.~(\ref{Ao})] of the coherent state will be squeezed into an even narrower
distribution.  In the limit of zero number fluctuation,
\begin{equation}
\langle (\Delta n_{1})^2\rangle = \langle (\Delta n_{2})^2\rangle =0,
\label{d2C} \end{equation}
the system becomes the Fock state
\begin{equation}
|F\rangle = \frac{a^{\dagger N/2}_{1} a^{\dagger N/2}_{2}}{(N/2)!}|0\rangle,
\end{equation}
which is clearly fragmented, since it is made up of two independent
condensates.  This fragmentation shows up clearly in the single particle
density matrix,
\begin{eqnarray}
  \langle a^{\dagger}_{\mu}a^{}_{\nu}\rangle &=& \frac{N}{2}\left(
  \begin{array}{cc} 1& 0\\
  0 & 1\end{array}\right)\\
&=&  \frac{N}{2} \left( \begin{array}{c} 1\\ 0 \end{array}\right)
  \left( \begin{array}{cc} 1 & 0 \end{array}\right) +
  \frac{N}{2} \left( \begin{array}{c} 0\\ 1 \end{array}\right)
  \left( \begin{array}{cc} 0 & 1 \end{array}\right),
\end{eqnarray}
which has two macroscopic eigenvalues, corresponding to independent
condensation in each well.  The evolution from the coherent state $|C\rangle$
to the Fock state $|F\rangle$ can be captured by the family of states
\begin{equation}
 \Psi_{\ell}(\sigma)  = \frac{e^{-\ell^2/\sigma^2}}{(\pi \sigma^2/2)^{1/4}}.
 \label{Fockfamily}
 \end{equation}
As $1/\sigma^2$ varies from $1/N$ to much greater than unity, the initial
coherent state $|C\rangle$ becomes more and more Fock-like.  Indeed, exact
numerical solution of Eq.~(\ref{Sch}) shows that Eq.~(\ref{Fockfamily}) is an
accurate description of the evolution from a coherent state towards a Fock
state.  During the collapsing process when the wavefunction $\Psi_{\ell}$
still extends over many number states (but fewer than $(N/2)^{-1/2}$), one can
take the continuum limit of Eq.~(\ref{Sch}), which reduces to the equation for
a particle in a harmonic oscillator potential.  From its Gaussian ground state
wavefunction, we extract $\sigma^{-2} = (2/N)(1 + UN/t)^{1/2}$.

Using this continuum approximation, we calculate the off-diagonal matrix element,
 \begin{eqnarray}
  \langle a^{\dagger}_{1} a^{}_{2}\rangle &=& \sum_{\ell} \sqrt{ (N/2 -
 \ell)(N/2 + \ell +1)} \Psi_{\ell +1}\Psi_{\ell} \\
 &\approx& (N/2) e^{-1/(2\sigma^2)},
\end{eqnarray}
which leads to a single particle density matrix
\begin{equation}
 \langle a^{\dagger}_{\mu}a^{}_{\nu}\rangle = \frac{N}{2}\left(
 \begin{array}{cc} 1 & e^{-1/2\sigma^2} \\ e^{-1/2\sigma^2} & 1
\end{array}\right).
\label{phasecoh}
\end{equation}
The eigenvalues are $\lambda_{1} = \frac{N}{2} (1 + e^{- 1/\sigma^2})$, and
 $\lambda_{2} = \frac{N}{2} (1 - e^{- 1/\sigma^2})$.   The relative number
fluctuations are
\begin{equation}
 \langle (\Delta n_{1})^2 \rangle
  = \sigma^2/2 .
 \label{d2F}
\end{equation}
As $1/\sigma^2$ varies from $1/N$ to a number much larger than unity, the
eigenvalues $(\lambda_{1}, \lambda_{2})$ varies from $(N, 0)$ to $(N/2, N/2)$,
and $ \sqrt{\langle (\Delta n_{1})^2 \rangle}$ varies from $\sqrt{N}$ to 0.

This transition of the coherent state into a Fock
state, described by the family Eq.~(\ref{Fockfamily}), is due to increasing
phase fluctuations, as discussed in sec.~\ref{dwmf}.  The phase fluctuation
effect can be seen by writing Eq.~(\ref{Fockfamily}) in terms of phase states,
\begin{eqnarray}
 |\phi\rangle &=&\frac{1}{\sqrt{ 2^{N} N!}} \left(  e^{i\phi/2}
 a^{\dagger}_{1} + e^{-i\phi/2} a^{\dagger}_{2}\right)^{N}|0\rangle
 \nonumber \\&=&
\sum_{\ell} e^{i\phi\ell}\Psi^{(0)}_{\ell}  |\ell\rangle,
\end{eqnarray}
 where $\Psi_{\ell}^{(0)}$ are the coefficients given by Eq.(\ref{Ao}).  The
family of q.(\ref{Fockfamily}) then becomes
\begin{equation}
 | \Psi(\sigma) \rangle  = \frac{  (\pi
 \overline{\sigma}^2)^{1/2}}{(\sigma^2/N)^{1/4}}
   \int ^{\pi}_{-\pi}  \frac{ {\rm d} \phi}{2\pi} e^{-\overline{\sigma}^2
 \phi^{2}/4} |\phi \rangle
 \label{phaseaverage}
 \end{equation}
where $\overline{\sigma}^{-2} = \sigma^{-2} - N^{-1}$.  As $\sigma^{-2}$
varies from $\sigma^{-2}\sim N^{-1}$ (coherent state) to $\sigma^{-2}>>1$, the
Gaussian in Eq.~(\ref{phaseaverage}) changes from a $\delta$-function,
$\delta(\phi),$ to a uniform distribution, thereby driving the coherent state
towards a Fock state.

    {\em Attractive Interaction:} When $U<0$, the potential energy
$U(n_{1}-n_{2})^2/4$ in Eq.~(\ref{U}) favors a large number difference between
the two wells, in particular, the states $|\ell =N/2\rangle = |N,0\rangle$ and
$|\ell = -N/2\rangle = |0, N\rangle$.  It therefore acts in the opposite
direction as hopping, which favors a Gaussian distribution of number states
around $|\ell=0\rangle$.  The effect of interaction is then to split the
Gaussian peak of the coherence state Eq.~(\ref{Ao}) into two peaks, a process
which can be described by the family of states \cite{HoCio},
\begin{equation}
     \Psi_{\ell}(a) = C \left(e^{-(\ell - a)^2/2\sigma'^2} + e^{-(\ell +
    a)^2/2\sigma'^2} \right)
\label{catfamily}
\end{equation}
where $2a$ is the separation between the peaks, $\sigma'$ is the width of
the peaks, and $C$ is the normalization constant.  As $a$ varies from 0 toward
$N/2$, and $\sigma'$ shrinks at the same time from $1/\sqrt{N}$ to 0, the
state $|\Psi(a)\rangle = \sum_{\ell} \Psi_{\ell} |\ell\rangle$ evolves from
the coherent state $|C\rangle$ to a Schr\"odinger-cat state
\begin{equation}
 |Cat\rangle = \frac{1}{\sqrt{2}} \left( |N,0\rangle + |0, N\rangle\right).
\label{cat} \end{equation}
The Schr\"odinger-cat state is fragmented in the sense that its single
particle density matrix has two large eigenvalues,
\begin{equation}
  \langle a^{\dagger}_{\mu}a^{}_{\nu}\rangle = \frac{N}{2}\left(
  \begin{array}{cc} 1 & 0 \\ 0 & 1
  \end{array}\right),
\end{equation}
identical to that of Fock state.  On the other hand, contrary to the Fock
state, it has huge number fluctuations,
\begin{equation}
 \langle (\Delta n_{1})^2 \rangle =   \langle (\Delta n_{2})^2 \rangle =
  N^2/4.
\end{equation}

    Details of how $a$ and $\sigma'$ depend on the ratio $U/t$ are found in
\cite{HoCio}, where it is also shown that the family Eq.~(\ref{catfamily})
accurately represents the numerical solution of Eq.~(\ref{Sch}).  This
double-well example brings out the important point that fragmented condensates
cannot be characterized by the single particle density matrix alone.  Higher
order correlation functions such as number fluctuations are needed.  This
example also shows how a coherent state can be brought into a Fock state (or
Schr\"odinger-cat state) through the phase (or number) fluctuations caused by
repulsive (or attractive) interaction.

\subsection{Spin-1 Bose gas}

    The spin-1 Bose gas, which is only marginally more complicated, provides
an excellent illustration of the role of symmetry in condensate fragmentation.
Here the degeneracies that give rise to fragmentation are due to a symmetry:
rotational invariance in spin space.  We will see that in the presence of
local antiferromagnetic interactions all low energy singly condensed states
break this symmetry.  The true ground state, which is a quantum superposition
of all members of this degenerate manifold, is a fragmented condensate.

    We consider a spin-1 Bose gas with $2N$ particles in the single mode
approximation, where each spin state has the same spatial wavefunction.
Several recent experiments have been carried out in this limit
\cite{sengstock,hirano,Chapman} (note that several of these experimental also explored regimes where the single mode approximation was not valid \cite{Saito}).  For conceptual clarity we consider the
extreme case where the trap is so deep that all particles reside in the lowest
harmonic state, trivially leading to the single mode limit.  In this extreme
situation there are no particle fluctuations: the only relevant interaction is
scattering between different spin states.  We denote the creation operator of
a boson with spin projection $\mu$ = (1,0,-1) in the lowest harmonic state by
$a^{\dagger}_{\mu}$; the number of bosons with spin $\mu$ is $N_{\mu}=
a^{\dagger}_{\mu} a^{}_{\mu}$.  The most general form of the scattering Hamiltonian which is rotational invariant in spin space is \cite{Ho}
\begin{equation}
  H= c{\bf S}^2
 \label{spin1H}
\end{equation}
where ${\bf S} = \sum_{\mu\nu} a^{\dagger}_{\mu} {\bf S}_{\mu\nu}
a^{}_{\nu}$, with $S^{i}_{\mu\nu}$ the spin-1 matrices, ($i=x,y,z$), and $c$
the interaction constant.  We consider the case $c>0$.  More complicated
Hamiltonians, with more interaction terms, are allowed for atoms with higher
spins \cite{HoLan,Ueda}.  The variety of fragmented states proliferate rapidly as
the atomic spin increases.

    To illustrate how condensate fragmentation is affected by external
perturbations, we include the linear Zeeman effect, with Hamiltonian,
\begin{equation}\label{magt}
  H_{Z}= - p S_{z} = -p(N_{1} - N_{-1}),
\end{equation}
where $p$ is the Zeeman energy proportional to the external magnetic field
$B$.  To explain current experiments one also needs to include the quadratic
Zeeman effect:  the atomic energy levels of atoms are not linear in $B$ due to
hyperfine interaction between electron spins and nuclear spins.  We ignore
these nonlinearities as they are irrelevant for describing fragmentation.

    For later discussion, we also include a term of the form
\begin{equation}
  H_{G}^{}= \epsilon (a^{\dagger}_{1} a_{-1}^{} + a^{\dagger}_{-1}a^{}_{1}),
\label{HG}\end{equation}
which mixes spin states $1$ and $-1$, where $\epsilon$ is a constant.
Terms of this form can be generated by magnetic field gradients
\cite{Ho1999a}, in which case $\epsilon$ is proportional to the square of the
field gradient.  Since both $H$ and $H_{Z}$ conserve $S_{z}$, the density
matrix $\langle a^{\dagger}_{\mu}a^{}_{\nu}\rangle$ for the ground state is
diagonal in the presence of these terms alone, and the system is generally
fragmented unless the density matrix happens to have only one macroscopic
eigenvalue.  The effect of $H_{G}$ is to mix the 1 and -1 states, bringing the
system into coherence, similar to the role of the tunneling term in the double
well system.

\subsubsection{Mean-field approach}

    Like the double well problem in the previous section, the Hamiltonian in
Eq.~(\ref{spin1H}) is exactly soluble.  As previously, we first consider the
mean-field solution so that we can relate the fragmented condensate to a
linear combination of singly condensed states.

    The general form of a singly condensed spinor Bose condensate is
\begin{equation}
  |\zeta \rangle_N = \frac{1}{\sqrt{N!}}\left(\sum_{\mu} \zeta_{\mu}
  a^{\dagger}_{\mu} \right)^{N}|0\rangle, \,\,\,\,\,\,\,\,  \sum_{\mu}|
   \zeta^{}_{\mu}|^{2}=1.
\label{mf}
\end{equation}
Using the fact that $a^{}_{\beta}a^{}_{\nu}|\zeta\rangle_{N}^{} =
\sqrt{N(N-1)}\zeta_{\beta}\zeta_{\nu}|\zeta\rangle^{}_{N-2}$, it is easy to
see that the number fluctuations are  $\langle N^{2}_{\nu} \rangle - \langle
N_{\mu}^{}\rangle^2 = \langle N_{\nu}^{}\rangle = N|\zeta_{\mu}^{}|^2$.
Writing the Hamiltonian as,
\begin{equation}
  H = {\bf S}_{\mu\nu}\cdot {\bf S}^{}_{\alpha\beta}\, a^{\dagger}_{\mu}
  a^{\dagger}_{\alpha} a^{}_{\beta}a^{}_{\nu} + 2\hat{N},
\end{equation}
we have, up to terms ${\cal O}(N^{-1})$,
\begin{equation}
  \langle H+H_{Z}^{}\rangle_{|\zeta\rangle_{N}} = c\langle {\bf S}\rangle^2 -
   p \langle S_{z}^{}\rangle + 2c N
\label{mean1}
\end{equation}
where $\langle {\bf S}\rangle = N\sum_{\mu,\nu}\zeta^{\ast}_{\mu}{\bf
S}_{\mu\nu} \zeta_{\nu}^{}$.  It is easy to see that Eq.(\ref{mean1}) is
minimized by
\begin{equation}
 \zeta^{T}_{\phi} = e^{i\gamma} \left( e^{-i\phi} \sqrt{\frac{N_{1}}{N}}, 0,
  e^{i\phi} \sqrt{\frac{N_{-1}}{N}}\right),
\label{meanzeta}
\end{equation}
where, as before, $T$ stands for transpose, $\phi$ is the relative phase
between the 1 and $-1$ component, $N_{1} + N_{-1}=N$, and the difference $M=
N_{1} - N_{-1} = \langle S_{z}\rangle$ is given by the nearest integer to
$p/2c$.  Note that $\langle S_{x}\rangle = \langle S_{y}\rangle=0$, and the
entire family $\{ \zeta_{\phi} \}$ is degenerate, reflecting the invariance of
Eq.~(\ref{mean1}) under spin rotation about the $z$-axis.  In the absence of a
magnetic field, $p\rightarrow 0$, we have $\zeta^{T}_{\phi} \rightarrow
e^{i\gamma} \left( e^{-i\phi}, 0, e^{i\phi}\right)/\sqrt{2}$.  However, since
Eq.~(\ref{mean1}) becomes fully rotationally invariant at zero field, any
arbitrary rotation of the state $\left( 1, 0, 1\right)/\sqrt{2}$ is also an
optimal spinor condensate.  The entire degenerate family (referred to as the
``polar" family in literature) is given by
\begin{equation}
 \zeta^{(0)}_{\phi} = e^{i\gamma} \left( \begin{array}{c} -
 \frac{1}{\sqrt{2}}e^{-i\phi} {\rm sin}\theta \\
 {\rm cos}\theta \\  \frac{1}{\sqrt{2}}e^{i\phi} {\rm sin}\theta
 \end{array}\right),
\label{polarfamily}
\end{equation}
with energy
\begin{equation}
 \langle H\rangle_{\rm polar} = 2cN.
\label{polarenergy}
\end{equation}

    Before discussing the exact ground states, we introduce the ``Cartesian"
operators
\begin{equation}
 A_{x} = -\frac{a_{1} + a_{-1}}{\sqrt{2}}, \,\,\,\,\, A_{y} =  \frac{a_{1} -
 a_{-1}}{i \sqrt{2}},
 \,\,\,\,\,  A_{z} = a_{0}.
\end{equation}
The important properties of these operators is that that under a spin
rotation $a_{\mu} \to U_{\mu\nu}a_{\nu}$, where $U= {\rm
exp}(-i\vec{\theta}\cdot {\bf S})$, ${\bf A}$ rotates like a Cartesian vector,
i.e.  $A_{i} \to R(\vec{\theta})_{ij} A_{j}$, where $R_{ij}(\vec{\theta})$ is
a rotational matrix in Cartesian space ($xyz$).  It is easily verified that
$[A_{i}^{}, A^{\dagger}_{j} ] = \delta_{ij}$, and
\begin{equation}
 {\bf S}= -i {\bf A}^{\dagger}\times {\bf A}, \,\,\,\,\,\,\,\,  N= {\bf
 A}^{\dagger} \cdot {\bf A},
\end{equation}
\begin{equation}
 H = c \left[ N^2 -\Theta^{\dagger }\Theta^{} \right]  ,
 \,\,\,\,\,\,\,\,\,\,\, \Theta = {\bf A}^{2} .
\end{equation}
and the operator that creates a singlet pair is,
\begin{equation}
 \Theta^{\dagger} = {\bf A}^{\dagger 2}= -2a^{\dagger}_{1}a^{\dagger}_{-1} +
 a^{\dagger 2}_{0}.
\label{Theta}
\end{equation}
The mean-field state Eq.~(\ref{mf}) can now be written as
\begin{equation}
 |\vec{\alpha} \rangle = \left(  \vec{\alpha}\cdot {\bf
 A}^{\dagger}\right)^{N}|0\rangle /\sqrt{N!}, \,\,\,\,\,\,\,\,
 \vec{\alpha}^{\ast}\cdot \vec{\alpha}=1.
\label{alpha}
\end{equation}
which has average spin
\begin{equation}
  \langle {\bf S} \rangle = -i  N \vec{\alpha}^{\ast} \times \vec{\alpha}
\end{equation}
The optimal mean-field state is determined by minimizing
\begin{equation}
 \langle \vec{\alpha}|H|\vec{\alpha}\rangle = c N^2 \left[ 1 -
 |\vec{\alpha}\cdot\vec{\alpha}|^2\right]
 - i p N \hat{\bf z} \cdot \vec{\alpha}^{\ast}\times \vec{\alpha}.
\end{equation}
In zero field ($p=0$), the optimal mean-field state is $\vec{\alpha}=
\hat{\bf n}$, where $\hat{\bf n}$ is a {\em real} unit vector.  The polar
family mentioned above can now be conveniently represented by all possible
directions of $\hat{\bf n}$, and the polar state in Eq.~(\ref{polarfamily})
corresponds to
\begin{equation}
 |\hat{\bf n}\rangle = \frac{ (\hat{\bf n}\cdot {\bf
 A}^{\dagger})^{N}}{\sqrt{N!}}|0\rangle .
\label{polar}
\end{equation}

\subsubsection{Exact ground state in a uniform magnetic field:}

    In the absence of a magnetic field, $H=c{\bf S}^2$.  Since $H$ is
proportional to the angular momentum operator, the ground state for $C>0$
(with an even number of bosons $N$) is a singlet.  The singlet state of a
single mode spin-1 Bose gas is unique \cite{HoLan}, and hence any manifestly
rotationally invariant state must be the ground state.  One can therefore
obtain the ground state by forming singlet pairs, which for an even number of
particles yields \cite{Koashi,Law1998c,Ho1999a}
\begin{equation}
 |S=0\rangle \propto({\bf A}^{\dagger }\cdot{\bf A}^{\dagger})^{N/2}|0\rangle.
\end{equation}
Since this state is rotationally invariant, its single particle density
matrix $\langle a^{\dagger}_{\mu}a^{}_{\nu}\rangle$ is proportional to
the identity matrix (Shur's theorem).  With the constraint $N = \sum_{\mu}
a^{\dagger}_{\mu}a^{}_{\mu}$, we then have
\begin{equation}
 \langle a^{\dagger}_{\mu}a^{}_{\nu}\rangle_{|S=o\rangle}=
\frac{N}{3}\delta_{\mu\nu}
\end{equation}
The condensate is therefore fragmented into three large pieces.  The
energy of the ground state is exactly zero, whereas that of the mean-field
state is $2cN$ (as shown in Eq.~(\ref{polarenergy})).

    To relate the exact ground to the optimal mean-field state (the polar
family $\{ |\hat{\bf n}\rangle \} $), we note that
\begin{equation}
 |S=0\rangle  \propto \int \frac{{\rm d}\hat{\bf n}}{4\pi} \left( \hat{\bf
 n}\cdot {\bf A}^{\dagger}\right)^{N}|0\rangle,
\label{spin1average}
\end{equation}
i.e., the exact ground state is an average over the family of optimal mean
field states.  This is the analog of the symmetry averaging relation in
Eq.~(\ref{F=C}).  This averaging process represents the quantum fluctuations
within the family of degenerate mean-field states; the effect of these
fluctuations is to reduce the mean-field energy from $2cN$ to zero.

    Next we consider the case of non-zero magnetic field $p\neq 0$. the
Hamiltonian is $H+ H_{Z} = c{\bf S}^2 - pS_{z}$.  The ground state is
$|S,S_{z}=S\rangle$ or simply denoted as $|S,S\rangle$, where $S$ is an
integer closest to $p/2c$.  The state $|S,S\rangle$ can be easily obtained by
changing a singlet pair into a triplet pair, and we have
\begin{equation}
 |S,S\rangle = {\cal D} a^{\dagger S}_{1} \left({\bf A}^{\dagger 2}\right)^{
 (N-S)/2}|0\rangle,
\label{SS}
\end{equation}
where ${\cal D}$ is a normalization constant, and Bose statistics require
$S$ to be even if $N$ is even.  Since the Hamiltonian $H+H_{Z}$ conserves
$S_{z}$, the single particle density matrix remains diagonal, $ \langle
a^{\dagger}_{\mu}a^{}_{\nu}\rangle = N_{\mu} \delta_{\mu\nu}$.  As shown in
\cite{Koashi,Ho1999a}, the eigenvalues $N_{\mu}$ have a very interesting behavior as
a function of $S$, namely
 \begin{equation}
 N_{1} = \frac{N(S+1) + S(S+2)}{2S+3},
\label{N1} \end{equation}
\begin{equation}
  N_{-1} = \frac{(N-S)(S+1)}{2S+3}, \,\,\,\,\,\,\,  N_{0}= \frac{N-S}{2S+3}.
\label{N0}
\end{equation}
If $S$ is of order 1, all three eigenvalues, $N_{\pm 1}, N_{0},$ are
macroscopic.  On the other hand, if $S$ becomes macroscopic, (i.e., $S/N$ is
less than but of order 1), both $N_{1}$ and $N_{-1}$ remain macroscopic (with
$N_{1}\rightarrow (N+S)/2$ and $N_{-1}\rightarrow (N-S)/2$) while $N_{0}$
becomes of order unity.  This means as $S$ increases, the $S_{z}=0$ component
is completely depleted.  Although the system is still fragmented, the number
of fragmented pieces is reduced from three to two, even for a tiny spin
polarization.  Further analysis \cite{Ho1999a} shows that as one increases
$S$, the fluctuations, $\langle \Delta N_1^2\rangle=\langle
N_1^2\rangle-\langle N_1\rangle^2$, drop rapidly.

    The reduction in the number of fragmented pieces mirrors the previously
discussed reduction in the size of the space of degenerate mean-field states.
In the absence of a magnetic field, polar states with arbitrary $\bf \hat n$
are degenerate, while a magnetic field in the $\bf \hat z$ direction favors
those with $\bf \hat n$ pointing in the x-y plane.

    To connect the exact polarized states to the mean-field states, we use the
relation ${\bf A}^{\dagger 2} = -2a^{\dagger}_{1} a^{\dagger}_{-1} +
a^{\dagger 2}_{0}$, Eq.~(\ref{Theta}), to write
\begin{equation}
 ({\bf A}^{\dagger 2})^{(N-S)/2}|0\rangle  = \sum_{p=0}^{(N-S)/2} D_{p}|p,
 N-S-2p, p\rangle
\label{sum}
\end{equation}
where the exact form of $D_p$ is not important
for our discussion.  If we now act on $({\bf A}^{\dagger
2})^{(N-S)/2}|0\rangle$ by $a_{1}^{\dagger}$, the coefficients of the {\em
large $p$} states in Eq.~(\ref{sum}) will be bosonically enhanced via
$a^{\dagger}_{1} |N_{1}\rangle = \sqrt{(N_{1} + 1)}|N_{1}+1\rangle$.  Further
acting on the same state by $a^{\dagger}_{1}$ eventually picks out the term
with largest $p$, which is $|(N+S)/2, 0, (N-S)/2\rangle$, in the sum
Eq.~(\ref{sum}).  This process is equivalent to replacing $a^{\dagger S}_{1}
(-2a^{\dagger}_{1} a^{\dagger}_{-1} + a^{\dagger 2}_{0})^{(N-S)/2}|0\rangle$
by $a^{\dagger S}_{1} (-2a^{\dagger}_{1}
a^{\dagger}_{-1})^{(N-S)/2}|0\rangle$.  We then have
\begin{eqnarray}
  |S,S\rangle&\propto& a^{\dagger S}_{1} ({\bf A}^{\dagger
  2})^{(N-S)/2}|0\rangle
  \approx   a^{\dagger S}_{1} (-2a^{\dagger}_{1} a^{\dagger}_{-1} )^{(N-S)/2}
  \nonumber \\
 &\propto &  a^{\dagger (N+S)/2}_{1}  a^{\dagger (N-S)/2}_{-1}|0\rangle.
\label{squeeze}
\end{eqnarray}
From our previous analysis of the two-well system it is clear that this
state is formed from an angular average of Eq.~(\ref{meanzeta}).  The field
gradient term (\ref{HG}) plays the role of tunneling in the two-well system
and can drive the system from a fragmented to coherent state.

    To complete the connection between these examples, we note that when the
occupation of the $S_z=0$ state is negligible, we can set $a_0=0$, and only
two spin states are required to describe the system.  Under these conditions,
the Hamiltonian for the spin 1 gas [Eqs.  (\ref{spin1H}),
(\ref{magt}),(\ref{HG})] reduces to the two-well Hamiltonian (\ref{2site})
with $U=4C$, and $t=-\epsilon$, and an asymmetry between the wells given by
$p$.  This analogy breaks down when $p\to0$ and the $S_z=0$ state becomes
occupied.

\subsection{ Fast Rotating Bose Gas:} \label{rotsec}

    In the previous cases of pseudospin-1/2 and spin-1 Bose gases, the number
of degenerate single particle states is of order unity, far fewer than the
number of bosons.  Bose gases with large amounts of angular momentum have just
the opposite behavior; the number of nearly degenerate states can be much
larger than the number of particles.  As we shall see, for relatively small
angular momentum, fragmentation similar to that in pseudospin-1/2 and spin-1
Bose gases can occur.  However, for very large angular momentum, it is
energetically more favorable for a repulsive Bose gas to organize itself into
a quantum Hall state, which is an extreme form of fragmentation in which all
traces of conventional condensation disappear.  Before discussing the
fragmentations of rotating Bose gas, we first discuss how the single particle
energy levels of a rotating Bose gas turn into Landau levels and achieve high
degeneracy as the angular momentum of the system increases.  For simplicity,
we limit the discussion to two dimensional systems.

    \subsubsection{ Lowest Landau levels (LLL) and the general properties of
many body wavefunctions in LLL:}

\begin{figure}
\includegraphics[width=\columnwidth]{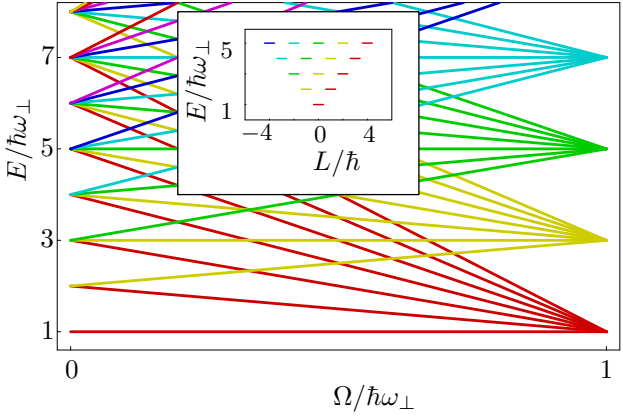}
\caption{(Color online) Energy eigenvalues of a two dimensional harmonic
oscillator as a function of angular velocity $\Omega$.  For clarity only seven
values of $m$ are shown, so that the infinite degeneracies at
$\Omega=\omega_\perp$ appear to be only seven-fold.  The inset shows the energy
states at $\Omega=0$ as a function of their angular momentum $L$.  (Colors:
each Landau level is colored with a distinct hue.)
\label{landau}}
\end{figure}

    The single particle Hamiltonian for a particle in the rotating harmonic
trap is
\begin{equation}
  h_{0}-\Omega L_{z}  = \frac{{\bf p}^{2}}{2M} + \frac{1}{2}
  M\omega_{\perp}^2r^2   - \Omega\hat{\bf z}\cdot {\bf r}\times {\bf p}.
\label{2Drot}
\end{equation}
where $\omega_{\perp}$ is the trap frequency, and $\Omega$ is the
rotational frequency of the trap.  Equation~(\ref{2Drot}) can be written as
\begin{equation}
 h_{0}-\Omega L_{z}  = \frac{({\bf p}- M\omega_{\perp} \hat{\bf z}\times {\bf
 r})^{2}}{2M} + \frac{1}{2} M(\omega_{\perp}^2
 -\Omega^{2}) r^2.
\label{new2Drot}
\end{equation}
If one rewrites ${\bf p} - M\omega_{\perp} \hat{\bf z} \times {\bf r}$ as
${\bf p} - e{\bf B}\times {\bf r}/2c$, with $eB/Mc=2\omega_{\perp}$, one sees
that Eq.~(\ref{new2Drot}) is identical to the Hamiltonian of an electron in a
magnetic field $B\hat{\bf z}$ in a harmonic potential with reduced frequency
$\omega^{2}_{\perp} - \Omega^{2}$.  
To diagonalize the single-particle Hamiltonian,
Eq.~(\ref{2Drot}),
we begin by noting that
the 2D simple harmonic oscillator $H_{0} = {\bf
p}^2/2M + M\omega^{2}_{\perp}r^2/2$ is diagonalized as $h_{0} = \hbar
\omega_{\perp}(a^{\dagger}_{x}a^{}_{x} + a^{\dagger}_{y}a^{}_{y}+1)$, where
$a_{x} = (d/\hbar)p_x-ix/d, a_{y} = (d/\hbar)p_y-iy/d$,
$d^2=\hbar/m\omega_\perp$, and $L_{z} = -i\hbar (a^{\dagger}_{x}a^{}_{y} -
a^{\dagger}_{y}a^{}_{x})$.  Defining $a_{\pm} = (a_{x} \pm i
a_{y})/\sqrt{2}$, we have
\begin{equation}
  L_{z} = -\hbar (a^{\dagger}_{+}a^{}_{+} - a^{\dagger}_{-} a^{}_{-}),
\end{equation}
and
\begin{equation}
 h_{0} -\Omega L_{z} = \hbar(\omega_{\perp}+ \Omega)a^{\dagger}_{+}a^{}_{+} +
 \hbar(\omega_{\perp}- \Omega)a^{\dagger}_{-}a^{}_{-} +\hbar\omega_\perp.
\end{equation}
The eigenstates are therefore
\begin{equation}
  |n,m\rangle = \frac{a^{\dagger n}_{+}}{\sqrt{n!}}\frac{a^{\dagger
  m}_{-}}{\sqrt{m!}}|0\rangle
\end{equation}
with eigenvalues
\begin{equation}
 E_{n,m} = \hbar(\omega_{\perp}+ \Omega)n +
 \hbar(\omega_{\perp}- \Omega)m+\hbar\omega_\perp,
\end{equation}
where $n, m= 0, 1, 2,\ldots$.  The energy levels, shown in
Fig.~\ref{landau}, are organized into Landau levels, labeled by $n$, separated
by $\delta E_1=\hbar(\omega_{\perp} + \Omega)$.  States within a given Landau level are
labeled by $m$ with spacing $\delta E_2=\hbar (\omega_{\perp}- \Omega)$.  At criticality,
$\Omega=\omega_{\perp}$, each Landau level becomes infinitely degenerate.

    The many-body Hamiltonian, including a contact interaction $g$, is
\begin{equation}
  H = \sum_{i=1}^{N} (h_{i}-\Omega L_{zi}^{} ) +  g \sum_{i>j}\delta({\bf
 r}_{i}-{\bf r}_{j})
\end{equation}
As $\Omega\to \omega_{\perp}$, mixing is strongest among the states in the
same Landau level.  To simplify matters, we consider only extremely weak
interactions, $gn << 2\hbar \Omega$.  This limit is naturally reached when
$\Omega$ is close to $\omega_\perp$, where the centrifugal potential largely
cancels the trapping potential and the cloud becomes large and dilute.  For
such weak interactions, only the lowest Landau level is populated.  
The eigenfunctions $u_{n=0, m}({\bf r})$ in the
lowest Landau level,
\begin{equation}
 u_{m}({\bf r}) = \frac{ (z/d)^{m}}{\sqrt{ \pi  d^2 m!} }
{e^{-|z|^2/2d^2}}\equiv \langle {\bf r}|m\rangle, \,\,\,\,\,\,\, z\equiv x+iy,
\end{equation}
are angular momentum eigenstates with $L=m$, and a spatial peak
at $\sqrt{\langle m| r^2 |m\rangle } =d \sqrt{2m+1} $.

    The many-body wavefunction for a systems of $N$ bosons is then
\begin{equation}
  \Psi({\bf r}_{1}, ..., {\bf r}_{N}) = \sum_{[m]} C_{[m]} u_{m_{1}}({\bf
  r}_{1})... u_{m_{N}}({\bf r}_{N}),
\end{equation}
where $\{m\}$ denotes the set of non-negative integers $(m_{1}, m_{2},
\dots m_{N})$.  Since, apart from the Gaussian factor, the single particle
wavefunctions are of the form $u_{m}\propto z^{m}$, the many-body wave
function is
\begin{equation}
 \Psi({\bf r}_{1}, ..., {\bf r}_{N}) = f(z_{1}, ..., z_{N}) e^{-
\sum_{i=1}^{N}|z_{i}|^2/2d^2}
\label{many}
\end{equation}
where $f$ is an analytic function which is symmetric in $z_{i}$.  In
particular, a Bose-condensed state corresponds to
\begin{equation}
 \Psi_{w}({\bf r}_{1}, ..., {\bf r}_{N}) = \prod_{i=1}^{N}  w({\bf r}_{i}),
 \,\,\,\,\,\,\,\,  w({\bf r}) = \sum_{m} \alpha_{m}u_{m}({\bf r}).
\label{w}
\end{equation}
If $\Psi$ is also an eigenstate of the total angular momentum $L_{z}= L$,
then $f$ in Eq.~(\ref{many}) must be a homogeneous symmetric polynominal of
$\{z_{i} \}$ of degree $L$.  Moreover, the single particle density matrix in
the angular momentum basis must be diagonal, i.e.,
\begin{equation}
 \langle a^{\dagger}_{m} a^{}_{n}\rangle= \delta_{mn}\langle a^{\dagger}_{m}
a^{}_{m}\rangle.
\end{equation}

    Thus a Bose condensed state (Eq.~(\ref{w})) that is also an angular
momentum eigenstate must be fully condensed into the single particles state
$u_{m}$, i.e., $w({\bf r})= u_{m}({\bf r})$.  The total angular momentum must
then be $L = mN\hbar$.  Such a coherent state, however, offers no flexibility
to redistribute particles to lower the energy, and is therefore unlikely to be
the ground state at fixed angular momentum or fixed rotational frequency
$\Omega$, except for certain special cases in which parameters of the system
are carefully tuned.  A general Bose condensed state, Eq.~(\ref{w}), would be
one where $w({\bf r})$ contains more than a single angular momentum state,
$u_{m}$.  Such states break rotational symmetry and are not 
angular momentum eigenstates.   Furthermore, due to
rotational symmetry, any rotation of $w$ ($w(r, \phi)\rightarrow w(r, \phi
+\theta)$) will have the same energy and angular momentum, $L= \langle
\Psi_{w}|L_{z} |\Psi_{w}\rangle$.  One can therefore form an angular momentum
eigenstate with angular momentum $L$ by forming the average
\begin{equation}
 \Phi_{L}({\bf r}_{1}, .. {\bf r}_{N})= \int {\rm d}\theta e^{iL\theta} \Psi(
\{ r_{i}, \phi_{i} + \theta\} ).
\label{angleave}
\end{equation}
Such averaging always lowers the energy of the system, and generically
results in a fragmented state.  Specific examples follow.  Note that with
small modification these arguments also apply in the general case when
interactions are not weak.

    \subsubsection{Ground state of the rotating Bose gas with attractive
interactions:}

    For the rotating Bose gas with attractive interaction, Wilkin, Gunn, and
Smith \cite{Wilkin1998a} pointed out the ground state with non-zero angular
momentum is one with all angular momentum carried by the center of mass, and
has the form
\begin{equation}
 \Psi({\bf r}_{1}, .. {\bf r}_{N}) =  K  (Z/d)^{L} e^{-
\sum_{i=1}^{N}|z_{i}|^2/2d^2},
\end{equation}
where $Z$ is the center of mass and $K$ is a normalization constant,
\begin{equation}
 Z = \frac{1}{N}\sum_{i=1}^{N}z_{i}, \,\,\,\,\,\,\,\,
 K=\frac{1}{\sqrt{  L!  (\pi d^2)^N} }.
\end{equation}
It is straightforward to show that the single particle density matrix of
this state is
\begin{equation}
 \langle a^{\dagger}_{m}a^{}_{n} \rangle = \delta_{mn}  N \left(  \frac{L!
 (N-1)^{L-m}}{m! (L-m)! N^{L}}
\right).
\label{fragatt}
\end{equation}
The ground state is fragmented because Eq.~(\ref{fragatt}) has a
distribution of large eigenvalues.  Later, Pethick and Pitaevskii \cite{pp}
pointed out that this state is of the form 
\begin{equation}
 \Psi( {\bf r}_{1}, .. {\bf r}_{N}) =  \Psi_{\rm CM}({\bf R}) \phi_{rel}(\{
 \vec{\rho}_{i} \}),
\end{equation}
where $\Psi_{\rm CM}({\bf R}) = (Z/d)^{L}e^{-|Z|^2/2d^2}$ is the wavefunction
of the center of mass ${\bf R}= \sum_{i=1}^{N}{\bf r}_{i}/N$, and
$\phi_{rel}(\{ \vec{\rho}_{i} \}) = K e^{-
\sum_{i=1}^{N}|\vec{\rho}_{i}|^2/2d^2}$ is a product of single particle states,
where the particle coordinates are $\{ \vec{\rho}_{i} =
{\bf r}_{i} - {\bf R} \}$.
Given this structure, it is natural to refer to this state as 
being singly condensed in the center
of mass frame.
  Here, we show that this exact ground state can
also be written as a symmetry average of broken symmetry states of the form
Eq.~(\ref{angleave}), as in many of the previous examples.

    Attractive interactions favor particles clumping together.  In homogeneous
systems, such clumping leads to collapse.  In the fast rotating limit,
however, the analyticity of the wavefunction in the lowest Landau level and
the Gaussian factor impose strong constraints on the degree of localization
possible.  The most localized state is $u_{0}({\bf r})$, a Gaussian of a width
given by the trap length $d$.  A similar localized wave-packet at location $a
\equiv a_{x} + ia_{y}$ is
\begin{eqnarray}
 \phi_{a}({\bf r}) &=  &e^{(2a^{\ast} z - |z|^2 -|a|^2)/2d^2}/\sqrt{\pi}\,d,
 \\
 &= & e^{- |z-a|^2/2d^2} e^{(a^{\ast}z - a z^{\ast})/2d^2} /\sqrt{\pi} d,
\label{phia}
\end{eqnarray}
which gives
\begin{equation}
|\phi_{a}({\bf r})|^2 = e^{-|z-a|^2/2d^2} /\pi d^2.
\end{equation}

    A many-body coherent state formed from these single particle states is
\begin{equation}
 \Psi({\bf r}_{1}, ... {\bf r}_{N}) = \prod_{i=1}^{N} \phi_{a}({\bf r}_{i}).
\end{equation}
This state carries angular momentum $\hbar L$ $=\langle \Psi| L_{z}| \Psi
\rangle$ $= \langle \Psi| \int {\rm d} {\bf r} \psi^{\dagger}({\bf r}) {\bf
r}\times (\hbar \nabla/i) \psi({\bf r})|\Psi\rangle $, with
\begin{equation}
 L  = N \int {\rm d}{\bf r}
  |\phi_{a}({\bf r})|^2 \left( [r/d]^2-1\right) = N |a/d|^2.
\end{equation}
The single particle density matrix is
\begin{equation}
 \langle \psi^{\dagger}({\bf r'})\psi({\bf r}) \rangle = N
 \phi^{\ast}_{a}({\bf r'})  \phi^{}_{a}({\bf r}),
\end{equation}
in which
\begin{eqnarray}
    \langle a^{\dagger}_{m}a^{}_{n} \rangle &=& \pi d^2 N u^{\ast}_{m}({\bf
   a}) u^{}_{n}({\bf a})                \nonumber \\
  &=& N \left( \frac{L}{N}\right)^{m} \frac{ e^{-L/N}}{m!}.
\end{eqnarray}
From this degenerate set of coherent states we construct an eigenstate of
angular momentum by taking the superposition
\begin{eqnarray}
  \frac{1}{d^{L+2}N^L \sqrt{L!}}\int \frac{ {\rm d}^{2} a}{2\pi}
  e^{-|a|^{2}/2d^2} a^{L}  |a\rangle \nonumber \\
  = \frac{Z^{L} e^{-
 \sum_{i=1}^{N}|z_{i}|^2/2d^2}}{\sqrt{  L!  (\pi d^2)^N} },
 \label{rotatingaverage}
\end{eqnarray}
where $Z = \frac{1}{N}\sum_{i=1}^{N}z_{i}/d$ .

    \subsubsection{Fragmentation in a rotating Bose gas with repulsive
interactions:}

    Similar fragmentation is found in the repulsive case, which for $L\ll
N\hbar$ is described by a vortex lattice.  A general mean field state is
of the form Eq.~(\ref{w}), with $w({\bf r}) = f(z)e^{-|z|^2/2d^2}$,
\begin{equation}
  f(z) = \prod_{a}(z-a),
\end{equation}
where the zeroes $a$ are the locations of the vortices \cite{hollvort}.  The
angular
momentum carried by this state can be obtained by noting that within the
lowest Landau level,
\begin{equation}
 \langle L_{z} \rangle = \hbar \langle \left(  (r/d)^2 -1\right) \rangle,
\end{equation}
which increases as the density of vortices increases.  Following the
arguments which we used for the attractive case, we are once again lead to
a fragmented condensate (see \cite{liu} for numerical studies).

    The rotating Bose gas with repulsive interactions is, however, much richer
than that with attractive interactions, since the mean-field picture breaks
down in a fundamental way at large values of angular momentum.  One sees this
breakdown by first noting that with increase of the angular momentum, the
vortex density increases and the particle density decreases
\cite{hollvort,baymllvort,coresize}.  Eventually, the density of vortices is
comparable to the density of particles and one can significantly improve the
energy of the system by correlating the positions of the vortices with the
positions of the particles.  These correlations cannot be captured by any
simple manipulation of the mean-field states; the general description of the
system with large angular momentum is quite complicated \cite{hall,wg}.

    Despite this complexity, there is a limit in which we can find the exact
ground state.  Imagine that the system is rotating with $\Omega$ sufficiently
close to $\omega_\perp$ that the energy spacing within the lowest Landau level
may be treated as a perturbation.  One would find the ground state by first
minimizing the interaction energy in the lowest Landau level, and then
perturbatively including the level spacing.  For short ranged interactions the
energy is minimized by any state for which the wavefunction vanishes whenever
two particles come together.  Degenerate perturbation theory then says that
the lowest energy state of all these wavefunctions is the one with lowest
angular momentum.  In the lowest Landau level, the lowest angular momentum
bosonic wavefunction which vanishes when two particles touch is the $\nu=1/2$
Laughlin state,
\begin{equation}
 \Psi([{\bf r}]) = \prod_{i>j}(z_{i}-z_{j})^2 e^{-|z_{i}|^2/2d^2}.
\end{equation}
For an infinite system, one readily sees that
\begin{equation}
 \langle a^{\dagger}_{m} a^{}_{n}\rangle = (1/2) \delta_{mn}
\end{equation}
for all $m$.  Not only are there no eigenvalues of order $N$, but they are
all less than unity.  All traces of the conventional Bose condensation are
obliterated.

\section{Discussion}
\subsection{Salient features}

    The examples shown in Sec.~III share a number of common features.
Many of the fragmentation processes we have discussed, such as those in
Eqs.~(\ref{phaseaverage}), (\ref{spin1average}), and (\ref{rotatingaverage}),
can be described by a family of quantum states that are weighted averages of
broken symmetry states over the space of broken symmetry, typically of the
form
\begin{equation}
 |\Psi (\lambda) \rangle =  \int {\rm d} \chi W(\chi, \lambda)
  |\chi\rangle_{c}^{} \,\,\, ,
\label{evolution}
\end{equation}
where $|\chi\rangle_{c}$ is a coherent state with broken symmetry
parameter $\chi$ (e.g., a spin direction), $W$ is a distribution function in
the space of broken symmetry, and $\lambda$ is the parameter that controls the
fragmentation of the system.  If, as $\lambda$ changes, say, from $0$ to 1,
$W$ changes from a distribution sharply peaked at $\chi_{0}$ to a completely
uniform distribution, Eq.~(\ref{evolution}) will evolve from the coherent
state $\chi_{0}$ to a fragmented state.  Such changes in the distribution
function reflect a growing fluctuation about the initial coherent state
$\chi_{0}$ singled out by a tiny symmetry-breaking field in the coherent
regime.

    For example, in the case of repulsive Bose gas in a double well, $\chi$ is
the relative phase between the condensates in the two wells and $\chi_{0}=0$.
In the spin-1 Bose gas, $\chi$ is the vector $\hat{\bf n}$ and $\chi_{0}$ is a
direction normal to an infinitesimal external magnetic field.  In the rotating
attractive Bose gas, $\chi$ is the location in space of the coherent state,
and $\chi_{0}$ is the equilibrium location determined by, e.g., a weak
potential that breaks rotational symmetry.  Exactly how the fluctuations about
$\chi_{0}$ grow depends on the specific dynamics of the system.  Large
fluctuations about $\chi_{0}$ directly reflect the competition different
degenerate states for Bose condensation, the ultimate cause of fragmented
structure, as discussed in the Introduction.

    Another important feature of states of the type (\ref{evolution}) is that,
as a consequence of bosonic enhancement, the range of the control parameter
$\lambda$ over which the system switches from a coherent to a fragmented state
shrinks with particle number $N$.  For a repulsive Bose gas in a double well,
as discussed in Sec.~III.A.2, the transition from a fragmented to a coherent
takes place around $t/U> 1/N$ \cite{HoCio}.  A similar situation occurs in the
spin-1 Bose gas, where the singlet (fragmented) state gives way to a coherent
state for field gradient $G >1/N$ \cite{Ho1999a}.  Since the window for
fragmentation vanishes as $N$ increases, fragmented condensates will not be
realized in condensed matter systems, with $N\sim 10^{23}$ particles.  On the
other hand, in mesocopic systems like quantum gases with typically $N\sim
10^{6}$ particles, fragmented states can exist in parameter ranges accessible
to experiment.

    Although our discussions have mainly focussed on the class of states
Eq.~(\ref{evolution}), other fragmentation processes do not lead to states
that are most naturally expressed in this manner, e.g., the Schr\"odinger-cat
family in Eq.~(\ref{catfamily}).  One can have distinct fragmented states,
such as the Fock state and the Schr\"odinger-cat state, with entirely
different properties but identical single particle density matrices.  Such
different states are therefore indistinguishable within the Penrose-Onsager
scheme.  To tell them apart, it is necessary to examine second order
correlation functions such as number fluctuations.  Even higher order
correlation functions are needed to distinguish the fragmented states of more
complicated systems, such as those of $F=2$ Bose gases of $^{87}$Rb, and the
recently realized spin-3 Bose gas of $^{52}$Cr \cite{pfau}.  In general, one
can expect a large variety of fragmented states, differing from each other by
high order correlation functions.

    The relative number fluctuations of a fragmented state (such as those in
the pseudo-spin 1/2 and spin-1 Bose gases discussed in Sec.  III) are a
measure of the stability of the state.  Huge fluctuations, such as in the
Schr\"odinger-cat state and the singlet state of the spin-1 Bose gas indicate
that the system is easily damaged by external perturbations.  Consider, for
example, a perturbation of the form $H' = \eta c^{\dagger} a_{1} + {\rm h.c.}$,
where $a_{1}$ is a boson in one of the wells in the double well example, or a
boson in the $S_{z}=1$ spin state of a spin-1 Bose gas, $c^{\dagger}$ adds a
particle in a different atomic state of the same boson (e.g., a plane wave
state in the background gas), and $\eta$ is a very small parameter.  The
perturbation $H'$ acting on the Fock state $a^{\dagger N/2}_{1}a^{\dagger
N/2}_{2}|0\rangle$ simply changes it to another Fock state, $c^{\dagger}
a^{\dagger N/2 -1}_{1}a^{\dagger N/2}_{2}|0\rangle$, which also has zero
number fluctuations, $\langle N_{1}^2 - \langle N_{1}\rangle^2\rangle =0$.  In
contrast, $H'$ acting on the Schr\"odinger-cat state $(a^{\dagger N}_{1}
+a^{\dagger N}_{2})|0\rangle$ collapses it into the state
$c^{\dagger}a^{\dagger N-1}_{1}|0\rangle$, immediately reducing the enormous
number fluctuations of the Schr\"odinger-cat state to zero.  More generally,
if the cat state $|\Psi_{\rm Cat}\rangle = |\Psi_{1}\rangle + |\Psi_{2}\rangle$ is
a sum of two Gaussians in number space, as shown in Eq.~(\ref{catfamily}),
with one Gaussian ($|\Psi_{1}\rangle$) peaked at $|N_{+}, N_{-}\rangle$ and
the other ($|\Psi_{2}\rangle$) peaked at $|N_{-}, N_{+}\rangle$, where $N_{+}
+ N_{-} = N$, and $N_{+}>>N_{-}$, the action of $H'$ on $|\Psi_{\rm Cat}\rangle$
considerably enhances $|\Psi_{1}\rangle$ and suppresses $|\Psi_{2}\rangle$.

    The large effect of the small perturbation $H'$ is due to bosonic
enhancement, which gives $H' |N_{+}, N_{-}\rangle = \sqrt{N_{+}}\,\,
c^{\dagger}|N_{+} -1, N_{-}\rangle $ and $H' |N_{-}, N_{+}\rangle =
\sqrt{N_{-}}\,\, c^{\dagger}|N_{-} -1, N_{+}\rangle $. Since $N_{+}>>N_{-}$,
the norm of $H' |N_{-}, N_{+}\rangle$ is considerably smaller than that of $H'
|N_{+}, N_{-}\rangle$.  As a result, $H'|\Psi_{\rm Cat}\rangle \approx
H'|\Psi_{1}\rangle \approx \sqrt{N_{+}} c^{\dagger}|N_{+} -1, N_{-}\rangle$,
which is no longer a Schr\"odinger-cat state.  In a spin-1 Bose gas, a similar
action changes the singlet state $(2a^{\dagger}_{1}a^{\dagger}_{-1} -
a^{\dagger 2}_{0})^{N/2}|0\rangle$ to one with much smaller relative number
fluctuations.  

    Since Schr\"odinger cat-like states can collapse into Fock states with the
slightest perturbation, and Fock states can easily be reassembled into a
single condensate by any small amount of tunneling between different
fragmented pieces, why should one bother with fragmented states?  Is
fragmentation relevant?  The point, as mentioned before, is that even though
fragmented states cannot be realized in macroscopic systems, the situation is
different for mesoscopic systems like trapped quantum gases.  The huge
reduction in particle number considerably relaxes the constraint of formation
of Fock states and Schr\"odinger cat-like states, and fragmented ground states
become realizable.  The phenomena of fragmentation becomes even richer
if it takes place in both real and spin space, such as with high spin bosons
in an optical lattices close to the Mott limit.  The combined effect of spin
degeneracy and spatial degeneracy (due to different isolated wells) produces a
great variety of quantum phases as the spin of boson increases.  In addition,
the singlet ground state of a spin-1 Bose gas can also be viewed as a
``resource" for singlets and may therefore be useful in developing quantum
teleportation protocols in optical lattices with spin-1 bosons.  Our
understanding of the properties of fragmented states, and in particular their
dynamics, is at such an early stage that it leaves open considerable room for
inventive ideas, which is where the excitement lies.

    We stress that we have described only the simplest types of fragmented
states.  Many other kinds which do not fall into the category we considered.
The $\nu=1/2$ Laughlin state discussed in Sec.~III.C.3 is an example at the
other extreme.  There are no large eigenvalues singled out in the density
matrix.  Instead, all the eigenvalues are identical and of order unity.  In
addition to level degeneracy, temperature effects can cause condensates to
fragment, a subject to which we turn in the next section.

\subsection{Role of temperature}\label{ft}

    Up to this point, the examples of fragmentation given in this paper
have focussed on the case where interaction-driven quantum fluctuations break
up the condensate.  It is intuitively clear that thermal fluctuations can play
a similar role.  A trivial example is given by a non-interacting Josephson
junction, governed by the Hamiltonian
\begin{equation}
  H = -t (a^\dagger b + b^\dagger a).
\end{equation}
This Hamiltonian is diagonal in the basis of symmetric and asymmetric
states, for which the creation operators are $(a^\dagger+b^\dagger)/\sqrt{2}$,
and $(a^\dagger-b^\dagger)/\sqrt{2}$, so the single particle density matrix is
diagonal in this basis.  At zero temperature only the symmetric state is
occupied -- a single condensate.  At very large temperature each of these
states is equally occupied so the system is fragmented.  For intermediate
temperatures, the occupation of each of these states is
\begin{eqnarray}
  n_{\stackrel{\scriptscriptstyle \rm s}{\scriptscriptstyle \rm a}}
 &=&\frac{\sum_{m=-N/2}^{N/2}(N/2 \pm m) e^{2t\beta m}}
 {\sum_{m=-N/2}^{N/2} e^{2t\beta m}}\\\nonumber
 &=& \frac{N}{2} \mp\left( \frac{N+1}{2} \coth(\beta t (N+1))
 -\frac{1}{2}\coth(\beta t)\right).
\end{eqnarray}
The upper signs denote the symmetric state (s) and the lower the
antisymmetric state (a).  The crossover between the singly condensed state at
$T=0$ and the fragmented state at $T\gg t$ is smooth.  Extensions of this
argument are relevant for spinor condensates in which one can, in principle,
have a hierarchy of transition temperatures, where the $k=0$ mode becomes
macroscopically occupied below some temperature $T_c$, but order in the
spin-channel does not occur until a lower temperature \cite{isoshima2000a}.

    A vortex lattice provides a qualitatively different example of how finite
temperature fragments a condensate.  Imagine a bucket of $^4$He rotating at
frequency $\Omega$.  The ground state of the system contains a triangular
array of vortices with $n_{\rm v}$ vortices per unit area.  At finite temperature
the vortex lattice is thermally excited, giving rise to a decay in the phase
correlations across the sample.  If we let $\psi(r)$ be the superfluid order
parameter coarse-grained on a scale large compared to the vortex spacing, then
according to \cite{baymvortex,baymllvort},  the correlation function $\langle
\psi^*(r^\prime)\psi(r^\prime)\rangle$ decays as $|r-r^\prime|^{-\eta}$, for large separations.
The exponent 
$\eta=1/(3\pi^2\rho r_{\rm v}\Lambda^2)$ is proportional to the ratio of the
distance between particles to the distance between vortices.  Here the
particle number density is $\rho$, the thermal wavelength is
$\Lambda=\sqrt{2\pi/mk_bT}$ and the distance between vortices is $r_{\rm v}$.  This
algebraic decay of correlations in real space corresponds to an algebraic
decay in momentum space.  For $\eta<3$, the occupation of the $k=0$ mode scale
as $N_0\sim L^{3-\eta}$, and as as $k\to0$ the occupations of the $k\neq 0$
modes scale as $N_k\sim k^{\eta-3}$.  Thus the number of macroscopically
occupied modes scales as $L^\eta$, and for carefully chosen $L$ and $\eta$ a
fragmented state can result.

    For a typical helium experiment $\eta< 10^{-8}$, and the depletion caused
by this effect is negligible.  Experiments on alkali gases, e.g.,
\cite{mitvort,jilavort}, create vortex lattices with $r_{\rm v}\sim 5 \mu$m and a
particle density $\rho\sim 10^{14}$ cm$^{-3}$, for which $\eta\sim 10^{-3}$,
also a minor correction.  Experiments with smaller vortex lattices in Paris
\cite{Madison2000a} have a similar vortex spacing $r_{\rm v}\sim2\mu$m, and density
$\rho\sim 10^{14}$ cm$^{-3}$, yielding a comparable value for $\eta$.
Although vortex lattices in current experiments are not thermally fragmented,
there does not appear to be any fundamental impediment to making $\eta$
larger.

\subsection{Other recent work}

    Not surprisingly, given the fascinating nature of fragmented states,
several recent papers (all cited in the relevant sections of this paper) have
been investigating models similar to those discussed here.  As a guide to the
reader, we briefly summarize these works, presenting them in roughly the same
order as they appear in the main text.

    The two-state system was explored by Nozi\`eres \cite{Nozieres1995a}, who
pointed out the existence of fragmentation in that system.  The generic
stability of a two-state fragmented condensate was considered by Rokhsar
\cite{Rokhsar1998a}.  One realization of this model would be to place atoms in
bonafide double-well potentials.  Due to mode mixing such a system is not
identical to a simple two-mode model, and several authors have explored more
realistic models.  Spekkens and Sipe \cite{spekkens} used a variational
approach to compare fragmented and singly condensed states in such potentials.
More recently, Streltsov and Cederbaum \cite{streltsov}, along with Moiseyev
\cite{cedermos} produced similar results through a multi-mode mean-field
theory.  The evolution of the ground state from a Fock state to a coherent
state and from a coherent state to a Schr\"odinger-cat state was studied by Ho
and Ciobanu \cite{HoCio}.  Experiments have recently produced a condensate
confined in a double-well trap \cite{dwexp}.

    The fragmented nature of the ground state of the spin-1 Bose gas was first
noted by Nozi\`eres and Saint James \cite{nsj}.  Later studies by Law, Pu, and
Bigelow \cite{Law1998c} investigated the ground state properties and the spin
dynamics in terms of the basis used in Eq.~(\ref{sum}).  Ho and Yip
\cite{Ho1999a} showed that the fragmented singlet state has huge fluctuations;
they also showed the relation between this fragmentation and spontaneous
symmetry breaking.  They explicitly showed how magnetic fields and field gradients drive
the system into a singly condensed state.  
Similar considerations were addressed by Koashi and Ueda~\cite{Koashi}.
Javanainen discussed
Issues involving the measurement of
the fragmented spin-1 ground state \cite{jav}.

    The rotating attractive gas was first studied by Wilkin et al.
\cite{Wilkin1998a}, who noted that the ground state was fragmented.  The
connection between this fragmentation and symmetry breaking was first
discussed by Pethick and Pitaevskii \cite{pp}.

    There have been extensive studies of the properties of the rotating
repulsive gas in the lowest Landau level -- many of which have focussed on the
structure of the order in the single particle density matrix.  These works
have mainly used a combination of exact diagonalization and variational
techniques.  A particularly relevant paper is the exact diagonalization study
of Liu et al.  \cite{liu}, which focusses on the symmetry-breaking nature of
the vortex states.  A related paper by Jackson et al.  \cite{jackson} compares
mean-field and exact wavefunctions at various values of the angular momentum.
The Goldstone mode associated with vortex nucleation was discussed by Ueda and Nakajima~\cite{UedaNakajima}.

    Fragmentation occurs in ``clumped" bosonic systems with attractive
interactions.  Ueda and Leggett used a two-mode approximation \cite
{uedaleggett} to study fragmentation and soliton formation in a one
dimensional attractive Bose gas.  A quite thorough comparison of mean-field
theory and exact diagonalzation is found in the articles by Kanamoto et al.
\cite{kanamoto}, as well as the closely related work of Kavoulakis
\cite{kav2003}.  Montina and Arecchi \cite{montina} use a Monte-Carlo scheme
to investigate the degree of fragmentation of this system.  In three
dimensions, Elgaroy and Pethick \cite{peth} showed that a harmonically trapped
gas of atoms with attractive interactions does not form a stable fragmented
state.

    Other systems with symmetry breaking and fragmentation include phase
separated two-component gases \cite{esry1999}, and rotating gases during a
phase-slip event \cite{muellerswallowtails}.  Boson ground states where the
condensate is broken into a macroscopic number of pieces include the Mott
insulator \cite{mott}, fractional quantum Hall states \cite{hall,wg}, and the
low dimensional Bose gas \cite{lowd,twoquasi}.  Aspects of the dynamics of
regaining phase coherence among condensates are discussed by Yi and Duan
\cite{dynamics}, a paper closely related to discussion of how the measurement
process is influenced by fragmentation \cite{Javmeasure,castindalibard}.

    Finally we mention the lecture notes produced by Castin and Herzog
for the 2000 Carg\`ese Summer School \cite{castin}, which lucidly introduces
fragmentation and analyzes the case of spin-1 bosons and of the
one-dimensional attractive gas.  They included an extended discussion of the role of symmetry breaking.

    \section{Summary}

    Bose condensation is remarkably robust.  It is therefore exciting to
search for zero-temperature bosonic states which are not condensed and to
understand the process that breaks up the condensate.  Fragmentation, where
the condensate breaks up into a few pieces, is the first step in this journey,
which eventually ends at strongly-correlated states possessing no trace of
condensation.

    We have described some of the canonical models which have fragmented
ground states, and extracted their properties.  We see that there is a rich
variety of fragmented ground states.  Often fragmentation is associated with
restoring a broken symmetry.  Sometimes it is accompanied by order in
higher-order correlation functions.  The one unifying feature appears to be a
combination of near-degeneracies and interactions.  The higher the degree of
degeneracy, the more fragmented the condensate may become.

\section{Acknowledgements}

    Parts of this research were performed at the Aspen Center for Physics and
the Kavli Institute for Theoretical Physics in Santa-Barbara.  The authors
acknowledge the following support:  EJM -- NSF Grant PHY-0456261 and the Sloan
foundation; TLH -- NSF Grant DMR-0426149 and NASA GRANT-NAG8-1765;  MU --
Grants-in-Aids for Scientfic Research (Grant No. 17071005) and a CREST program of the JST; and GB -- NSF Grant PHY03-55014
and PHY05-00914.


\begin{thebibliography}{99}

    \bibitem{huang} K. Huang, Statistical Mechanics (2nd Ed.), (Wiley, 1987).

\bibitem{einstein} A. Einstein, Ber. Berl. Akad. 261 (1924); 3 (1925).

    \bibitem{london} F. London, Phys.  Rev.  {\bf 54}, 947 (1938).

    \bibitem{bogoliubov} N. N. Bogoliubov, J. Phys.  (USSR) {\bf 11}, 23
(1947); reprinted in D. Pines, {\em The Many-Body Problem}, (W.  A. Benjamin,
New York, 1961).

    \bibitem{Penrose1956a} O. Penrose and L. Onsager, Phys.  Rev.  {\bf 104},
576 (1956).

    \bibitem{boseatoms} C. J. Pethick and H. Smith, Bose-Einstein Condensation
in Dilute Gases, (Cambridge Univ.  Press, Cambridge, 2002).

    \bibitem{nsj} P. Nozi\`eres and D. Saint James, J. Physique {\bf 43}, 1133
(1982).

    \bibitem{Nozieres1995a} P. Nozi\`eres, in {\em Bose-Einstein
Condensation}, edited by A. Griffin, D. W. Snoke, and S. Stringari (Cambridge
University Press, Cambridge, 1995).

    \bibitem{Wilkin1998a} N. Wilkin, J. Gunn, and R. Smith, Phys.  Rev.  Lett.
{\bf 80}, 2265 (1998).

    \bibitem{girardeau} M. Girardeau, Phys.  Fluids {\bf 5}, 1468 (1962).

 \bibitem{Koashi} M. Koashi and M. Ueda, Phys. Rev. Lett. {\bf 84}, 1066 (2000).

    \bibitem{Ho1999a} T.-L.  Ho and S. K. Yip, Phys.  Rev.  Lett.  {\bf 84},
4031 (2000).

    \bibitem{castin} Y. Castin and C. Herzog, ``Bose-Einstein condensates in
symmetry breaking states'', Comptes Rendus de l'Academie des Sciences de
Paris, tome 2, serie IV, 419 (2001).  Also available as {\tt
cond-mat/0012040}.

    \bibitem{peth} O. Elgaroy and C. J. Pethick, Phys.  Rev.  A {\bf 59}, 1711
(1999).

    \bibitem{Rokhsar1998a} D. S. Rokhsar, {\em Phase coherence and
``fragmented'' Bose condensates} (1998), {\tt cond-mat/9812260}.

    \bibitem{spekkens}R.  W. Spekkens and J. E. Sipe, Phys.  Rev.  A {\bf 59},
3868 (1999).

    \bibitem{jav} J. Javanainen, J. Phys.  B {\bf 33}, 5493 (2000).

    \bibitem{dukelsky} J. Dukelsky and P. Schuck, Phys.  Rev.  Lett.  {\bf
86}, 4207 (2001).

    \bibitem{jilaspinexp} C.J.  Myatt, E.A.  Burt, R.W.  Ghrist, E.A.  Cornell
and C.E.  Wieman, Phys.  Rev.  Lett.  {\bf 78}, 586 (1997); M.R. Matthews,
D.S.  Hall, D.S.  Jin, J.R.  Ensher, C. Wieman, E.A.  Cornell, F. Dalfovo, C.
Minniti, and S. Stringari, Phys.  Rev.  Lett.  {\bf 81}, 243 (1998); D. S.
Hall, M.R. Matthews, J.R.  Ensher, C.E.  Wieman, and E.A.  Cornell, Phys.
Rev.  Lett. 81, 1539 (1998); D.S.  Hall, M.R. Matthews, C.E.  Wieman, and E.A.
Cornell, Phys.  Rev.  Lett.  {\bf 81}, 1543 (1998); M.R. Matthews, B.P.
Anderson, P.C.  Haljan, D.S.  Hall, M.J. Holland, J.E Williams, C.E.  Wieman,
E.A.  Cornell, Phys.  Rev.  Lett.  {\bf 83}, 3358 (1999).

    \bibitem{spinexperiments} A. G\"orlitz, T.L.  Gustavson, A.E.  Leanhardt,
R. L\"ow,  A.P.  Chikkatur, S. Gupta, S. Inouye, D.E.  Pritchard, and W.
Ketterle, Phys.  Rev.  Lett.  {\bf 90}, 090401 (2003); D.M.  Stamper-Kurn,
H.-J.  Miesner, A.P.  Chikkatur, S. Inouye, J. Stenger, and W. Ketterle:
Phys.  Rev.  Lett.  {\bf 83}, 661 (1999).  H.-J.  Miesner, D.M.  Stamper-Kurn,
J. Stenger, S. Inouye, A.P.  Chikkatur, and W. Ketterle:  Phys.  Rev.  Lett.
{\bf 82}, 2228 (1999).  J. Stenger, S. Inouye, D.M.  Stamper-Kurn, H.-J.
Miesner, A.P.  Chikkatur, and W. Ketterle:  Nature {\bf 396}, 345 (1998).
D.M.  Stamper-Kurn, M.R. Andrews, A.P.  Chikkatur, S. Inouye, H.-J.  Miesner,
J. Stenger, and W. Ketterle:  Phys.  Rev.  Lett.  {\bf 80}, 2027 (1998); J. M.
Higbie, L. E. Sadler, S. Inouye, A. P. Chikkatur, S.R.  Leslie, K.L.  Moore,
V. Savalli, and D.M.  Stamper-Kurn.  Phys.  Rev.  Lett.  {\bf 95}, 050401
(2005); M.-S.  Chang, C.D.  Hamley, M.D. Barrett, J.A.  Sauer, K.M.  Fortier,
W. Zhang, L. You, and M.S. Chapman, Phys.  Rev.  Lett.  {\bf 92}, 140403
(2004); H. Schmaljohann, M. Erhard, J. Kronj\"ager, M. Kottke, S. van Staa, L.
Cacciapuoti, J.J.  Arlt, K. Bongs, and K. Sengstock, Phys.  Rev.  Lett.  {\bf
92}, 040402 (2004); and T. Kuwamoto, K. Araki, T. Eno, and T. Hirano, Phys.
Rev.  A {\bf 69}, 063604 (2004).

    \bibitem{1dexp} H. Moritz, T. St\"oferle, M. K\"ohl, and T. Esslinger,
Phys.  Rev.  Lett. {\bf 91}, 250402 (2003); T. St\"oferle, H. Moritz, C.
Schori, M. K\"ohl, and T. Esslinger, Phys.  Rev.  Lett. {\bf 92}, 130403
(2004).

    \bibitem{mott} M. Greiner, O. Mandel, T. Esslinger, T. W. Hansch, and I.
Bloch, Nature, {\bf 415}, 39 (2002); and M. P. A. Fisher, P. B. Weichmann, G.
Grinstein, and D. S. Fisher, Phys.  Rev.  B {\bf 40}, 546 (1989).

    \bibitem{hall} N.R.  Cooper and N.K.  Wilkin, Phys.  Rev.  B {\bf 60},
R16279 (1999); S. Viefers, T.H.  Hansson, and S.M.  Reimann, Phys.  Rev.  A
{\bf 62}, 053604 (2000); N.R.  Cooper, N.K.  Wilkin, and J.M.F.  Gunn, Rev.
Lett.  {\bf 87}, 120405 (2001); B. Paredes, P. Fedichev, J.I.  Cirac, and P.
Zoller, Rev.  Lett.  {\bf 87}, 010402 (2001); J.W.  Reijnders, F.J.M. van
Lankvelt, K. Schoutens, and N. Read, Phys.  Rev.  Lett.  {\bf 89}, 120401
(2002); B. Paredes, P. Zoller, and J.I.  Cirac, Phys.  Rev.  A {\bf 66},
033609 (2002);  N. Regnault, and Th.  Jolicoeur, Phys.  Rev.  Lett.  {\bf
91}, 030402 (2003); and T. Nakajima and M. Ueda, Phys. Rev. Lett. {\bf 91}, 
140401 (2003).

    \bibitem{wg} N.K.  Wilkin and J.M.F.  Gunn, Phys.  Rev.  Lett.  {\bf 84},
6 (2000).

   \bibitem{Penrose1951a} O. Penrose, Phil.  Mag. {\bf 42}, 1373 (1951).

    \bibitem{dubois} See, e.g., J.L. DuBois and H.R. Glyde, Phys.  Rev.  A
{\bf 68}, 033602 (2003).

    \bibitem{lowd} T. Papenbrock, Phys.  Rev.  A {\bf 67}, 041601(R) (2003);
D.M. Gangardt, and G.V.  Shlyapnikov, Phys.  Rev.  Lett.  {\bf 90}, 010401
(2003); M.D. Girardeau, E.M.  Wright, and J.M.  Triscari, Phys.  Rev.  A {\bf
63}, 033601 (2001); D.S.  Petrov, G.V.  Shlyapnikov, and J.T.M.  Walraven,
Phys.  Rev.  Lett.  {\bf 85}, 3745 (2000); and A. Lenard, J. Math.  Phys.
{\bf 7}, 1268 (1966).

    \bibitem{Javmeasure} J. Javanainen and S. M. Yoo, Phys.  Rev.  Lett.  {\bf
76}, 161 (1996).

    \bibitem{castindalibard} Y. Castin and J. Dalibard, Phys.  Rev.  A {\bf
50}, 4330 (1997).

    \bibitem{mitinterference} M. R. Andrews, C. G. Townsend, H.-J.  Miesner,
D. S. Durfee, D. M. Kurn, and W. Ketterle, Science {\bf 275}, 637 (1997); M.
R. Andrews, D. M. Kurn, H.-J.  Miesner, D. S. Durfee, C. G. Townsend, S.
Inouye, and W. Ketterle, Phys.  Rev.  Lett.  {\bf 79}, 553 (1997); {\em ibid.}
{\bf 80}, 2967 (1998).

    \bibitem{puzzle} The real puzzle is why each single shot measurement
yields an interference pattern of two phase-coherent condensates with an
apparent relative phase $\phi$.  The explanation, which is not contained in
Eq.(\ref{F=C}), lies in quantum measurement theory; the very act of
measurement actually establishes a phase between the two initially incoherent
condensates, a phenomenon that is a consequence of bosonic enhancement
\cite{castindalibard}.

    \bibitem{dwexp} M. Albiez, R. Gati, J. F\"olling, S. Hunsmann, M.
Cristiani, and M.K. Oberthaler, Phys.  Rev.  Lett.  {\bf 95}, 010402 (2005).

    \bibitem{uedaleggett} M. Ueda and A.J.  Leggett, Phys.  Rev.  Lett.  {\bf
83}, 1489 (1999).

    \bibitem{muellerswallowtails} E.J.  Mueller, Phys.  Rev.  A {\bf 66},
063603 (2002).

    \bibitem{lqm} G. Baym, {\em Lectures on Quantum Mechanics}, W. A.
Benjamin, Inc., New York, 1969.

    \bibitem{spintrans} By a suitable change of variables one can make the
hopping uniform, but at the cost of making the potential more complicated.
See G. Scharf, W. F. Wreszinski, and J. L. van Hemmen, J. Phys.  A:  Math.
Gen.  {\bf 20}, 4309 (1987).

    \bibitem{HoCio} T.-L.  Ho and C.V. Ciobanu, J. Low Temp.  Phys.  {\bf
125}, 257 (2004).

    \bibitem{sengstock} H. Schmaljohann, M. Erhard, J. Kronj\"ager, M. Kottke,
S. van Staa, L. Cacciapuoti, J.J. Arlt, K. Bongs, and K. Sengstock, Phys.
Rev.  Lett. {\bf 92}, 040402 (2004).

    \bibitem{hirano} T. Kuwamoto, K. Araki, T. Eno, and T. Hirano, Phys.  Rev.
A {\bf 69}, 063604 (2004).

    \bibitem{Chapman} W. Zhang, D.L. Zhou, M.-S.  Chang, M.S. Chapman, and
L. You, Phys.  Rev.  A {\bf 72}, 013602 (2005); and M.-S.  Chang, Q. Qin, W.
Zhang, L. You, M.S. Chapman, cond-mat/0509341 (2005).

  \bibitem{Saito} H. Saito and M. Ueda, Phys. Rev. A {\bf 72}, 023610 (2005).

    \bibitem{Ho} T.L.  Ho, Phys.  Rev.  Lett.  {\bf 81}, 742 (1998).

    \bibitem{HoLan} T.-L.  Ho and L. Yin, Phys.  Rev.  Lett.  {\bf 84}, 2302
(2000).

 \bibitem{Ueda} M. Ueda and M. Koashi, Phys. Rev. A {\bf 65}, 063602 (2002).


    \bibitem{Law1998c} C.K. Law, H. Pu, and N.P. Bigelow, Phys.  Rev.  Lett.
{\bf 81}, 5257 (1998).

    \bibitem{pp} C.J.  Pethick and L. Pitaevskii, Phys.  Rev.  A {\bf 62},
033609 (2000).

    \bibitem{hollvort} T.-L. Ho, Phys.  Rev.  Lett. {\bf 87}, 060403 (2001).

    \bibitem{liu} X.-J.  Liu, H. Hu, L. Chang, W. Zhang, S.-Q.  Li, and Y.-Z,
Wang, Phys.  Rev.  Lett.  {\bf 87}, 030404 (2001).

    \bibitem{baymllvort} G. Baym, Phys.  Rev.  A {\bf 69}, 043618 (2004).

    \bibitem{coresize} G. Baym and C.J.  Pethick, Phys.  Rev.  A {\bf 69},
043619 (2004).

    \bibitem{pfau} A. Griesmaier, J. Werner, S. Hensler, J. Stuhler, and T.
Pfau, Phys.  Rev.  Lett.  {\bf 94}, 160401 (2005); cond-mat/0503044; and A.
Griesmaier, J. Stuhler, and T. Pfau, cond-mat/0508423.

    \bibitem{isoshima2000a} T. Isoshima, T. Ohmi, and K. Machida, J. Phys.
Soc.  Jpn.  {\bf 69}, 3864 (2000).

    \bibitem{baymvortex} G. Baym, Phys.  Rev.  B {\bf 51}, 11697 (1995).

    \bibitem{mitvort} J. R. Abo-Shaeer, C. Raman, J. M. Vogels, W. Ketterle,
Science, {\bf 292}, 476 (2001).

    \bibitem{jilavort} V. Schweikhard, I. Coddington, P. Engels, V. P.
Mogendorff, and E. A. Cornell, Phys.\ Rev.\ Lett.\ {\bf 92}, 040404 (2004); I.
Coddington, P. C. Haljan, P. Engels, V. Schweikhard, S. Tung, and E. A.
Cornell, Phys.\ Rev.\ A {\bf 70}, 063607 (2004); and references therein.

    \bibitem{Madison2000a} K.W.  Madison, F. Chevy, W. Wohlleben, and J.
Dalibard, Phys.  Rev.  Lett.  {\bf 84}, 806 (2000).

    \bibitem{streltsov} A.I. Streltsov and L.S. Cederbaum, Phys.
Rev.  A {\bf 71}, 063612 (2005).

    \bibitem{cedermos} A.I.  Streltsov, L.S.  Cederbaum, and N. Moiseyev,
Phys.  Rev.  A {\bf 70}, 053607 (2004).

    \bibitem{jackson} A.D.  Jackson, G.M.  Kavoulakis, B. Mottelson, and S.M.
Reimann, Phys.  Rev.  Lett.  {\bf 86}, 945 (2001).

    \bibitem{UedaNakajima} M. Ueda and T. Nakajima, Phys. Rev. A {\bf 73}, 
    043603 (2006).


    \bibitem{kanamoto} R. Kanamoto, H. Saito, and M. Ueda, Phys.  Rev.  A {\bf
67}, 013608 (2003); {\bf 68}, 043619 (2003); and Phys.  Rev.  Lett.  {\bf 94},
090404 (2005).

    \bibitem{kav2003} G. M. Kavoulakis, Phys.  Rev.  A, {\bf 67}, 011601(R)
(2003); {\bf 69}, 023613 (2004).

    \bibitem{montina} A. Montina and F. T. Arecchi, Phys.  Rev.  A {\bf 71},
063615 (2005).

    \bibitem{esry1999} B. D. Esry, and C. H. Greene, Phys.  Rev.  A, {\bf 59},
1457 (1999).

    \bibitem{twoquasi} Y. Kagan, V.A.  Kashurnikov, A.V.  Krasavin, N.V.
Prokof'ev, and B.V.  Svistunov, Phys.  Rev.  A {\bf 61}, 043608 (2000); J.M.
Kosterlitz and D.J.  Thouless, J. Phys.  C {\bf 6} 1181 (1973); D.S.  Petrov,
M. Holzmann, and G.V.  Shlyapnikov, Phys.  Rev.  Lett.  {\bf 84}, 2551 (2000);
D.S.  Fisher and P. C. Hohenberg, Phys.  Rev.  B, {\bf 37}, 4936 (1988); V. N.
Popov, {\em Functional Integrals in Quantum Field Theory and Statistical
Physics}, (Reidel, Dordrecht, 1983).

    \bibitem{dynamics} W. Yi and L.-M.  Duan, Phys.  Rev.  A {\bf 71}, 043607
(2005).

\end{thebibliography}
\end{document}